\documentclass[aps, prd, twocolumn, showpacs, floatfix, letterpaper,nofootinbib, superscriptaddress]{revtex4-2}

\usepackage{blindtext}
\usepackage[newitem]{paralist}
\usepackage{graphicx}
\usepackage{epsfig}
\usepackage{bm}
\usepackage{amsfonts}
\usepackage{pgf}
\usepackage{color}
\usepackage[T1]{fontenc}
\usepackage[utf8]{inputenc}
\usepackage{graphicx}
\usepackage[english]{babel}
\usepackage{amsmath}
\usepackage{amssymb}
\usepackage{amsfonts}
\usepackage{makecell}
\usepackage{hyperref}
\usepackage[draft,deletedmarkup=xout]{changes}
\usepackage{mathtools}
\usepackage{xcolor}
\usepackage{multirow}
\usepackage{mhchem}
\definecolor{green}{rgb}{0.19,0.64,0.54}
\definecolor{blue}{rgb}{0,0,1}
\definecolor{reddish}{rgb}{0.65, 0.2, 0.2}
\definecolor{darkgreen}{rgb}{0.2,0.7,0.3}
\definecolor{darkblue}{rgb}{0.3,0.40,0.48}
\definecolor{gray}{rgb}{.8,.8,.8}

\hypersetup{,
	pdftitle={LoopsCMB},
  	pdfauthor={I.Yu. Rybak and L. Sousa},
    colorlinks=true,       
    linkcolor=darkblue,          
    citecolor=darkgreen,        
    filecolor=reddish,      
    urlcolor=blue,          
    linktocpage=true
}

\usepackage{wrapfig}
\usepackage{lipsum}
\newcommand{\be}{\begin{equation}}
\newcommand{\ee}{\end{equation}}
\newcommand{\bq}{\begin{eqnarray}}
\newcommand{\eq}{\end{eqnarray}}
\newcommand{\vv}[0]{{\bar v}}
\newcommand{\cc}[0]{{\tilde c}}
\newcommand{\kk}[0]{\mathbf{k}}

\usepackage{ulem}

\begin{document}

\title{CMB anisotropies generated by cosmic string loops}

\author{I. Yu. Rybak}
\email[]{Ivan.Rybak@astro.up.pt}
\affiliation{Centro de Astrof\'{\i}sica da Universidade do Porto, Rua das
Estrelas, 4150-762 Porto, Portugal}
\affiliation{Instituto de Astrof\'{\i}sica e Ci\^encias do Espa\c co,
CAUP, Rua das Estrelas, 4150-762 Porto, Portugal}

\author{L. Sousa}
\email{Lara.Sousa@astro.up.pt}
\affiliation{Centro de Astrof\'{\i}sica da Universidade do Porto, Rua das
Estrelas, 4150-762 Porto, Portugal}
\affiliation{Instituto de Astrof\'{\i}sica e Ci\^encias do Espa\c co,
CAUP, Rua das Estrelas, 4150-762 Porto, Portugal}

\begin{abstract}

We investigate the contribution of cosmic string loops to the Cosmic Microwave Background (CMB) anisotropies. This is done by extending the Unconnected Segment Model (USM) to include the contribution of the cosmic string loops created throughout the cosmological evolution of a cosmic string network to the stress-energy tensor. We then implement this extended USM in the publicly available CMBACT code and obtain the linear CDM power spectrum and the CMB angular power spectra generated by cosmic string loops. We find that the shape of the angular power spectra generated by loops is, in general, similar to that of long strings. However, there is generally an enhancement of the anisotropies on small angular scales. Vector modes produced by loops dominate over those produced by long strings for large multipole moments $\ell$. The contribution of loops to the CMB anisotropies generated by cosmic string networks may reach a level of $10\%$ for large loops but decreases as the size of loops decreases. This contribution may then be significant and, thus, this extension provides a more accurate prediction of the CMB anisotropies generated by cosmic string networks.
\end{abstract}

\date{\today}
\maketitle

\section{Introduction}

The Cosmic Microwave Background (CMB) has provided us with an accurate observational probe of several cosmological paradigms. The improved precision of measurements of the CMB anisotropies has led to stringent constraints of cosmological parameters, which translate into constraints on various early-universe scenarios. One such scenario is the production of topological defects networks in symmetry-breaking phase transitions in the early universe (see~\cite{HindmarshKibble, ShellardVilenkin, CopelandKibble} for a review). Although CMB observations are consistent with the inflationary paradigm~\cite{Fixsen:1996nj}, in which the perturbations are seeded in the very early universe, they still allow for a subdominant defect contribution. Current data limits the fractional contribution of line-like defects known as cosmic string to $1-2\%$ of the temperature anisotropies, which translates into a bound on cosmic string tension of $G \mu \lesssim 10^{-7}$~\cite{Planck2013,  LizarragaUrrestillaDaverioHindmarshKunz,LazanauShellard1, LazanauShellard, CharnockAvgoustidisCopelandMoss}. Despite this, cosmic strings may still contribute significantly to the B-mode polarization of the CMB and thus this signal may provide us with a relevant window to probe string-forming scenarios.

The derivation of accurate constraints on cosmic-string-forming scenarios requires a detailed prediction of the CMB anisotropies generated by a cosmic string network. Cosmic strings and other topological defects, however, source perturbations actively and thus the computation of their CMB signatures requires an understanding of how these perturbations are generated throughout cosmological history. This may be achieved by using numerical simulations \cite{FraisseRingevalSpergelBouchet, LazanauShellard1, LizarragaUrrestillaDaverioHindmarshKunz} of cosmic string networks, through analytic estimation \cite{BrandenbergerTurok,TraschenTurokBrandenberger} or by resorting to the publicly available CMBACT code~\cite{PogosianVachaspati}. This numerical tool, developed to compute the anisotropies generated by active sources, uses the Unconnected Segment Model~\cite{AlbrechtBattyeRobinson} to describe the stress-energy tensor of a cosmic string network. This framework, developed to describe standard cosmic string networks, is very versatile and was successfully extended to describe the CMB signatures of cosmic superstrings~\cite{CharnockAvgoustidisCopelandMoss},
superconducting strings~\cite{RybakAvgoustidisMartins},
and 2+1-dimensional topological defects known as domain walls~ \cite{SousaAvelino}. The USM was further extended in~\cite{AvgoustidisCopeland} to reduce computational time, which allowed for its use in Markov-Chain-Monte-Carlo analysis of the string parameter space.

The original USM only considers the contribution of long strings and does not include the loops that are copiously produced as a result of strings interactions. The loops' contribution to the CMB anisotropies is expected to be subdominant. However, a significant fraction of the string energy density is, at any time, in the form of loops. As a matter of fact it was shown in~\cite{TraschenTurokBrandenberger, Wu:1998mr} that loops indeed provide a significant contribution to the spectrum of perturbations generated by the cosmic string network, which may translate into a significant contribution to CMB anisotropies. Note that this contribution was not quantified as of yet (see however Refs.~\cite{BrandenbergerTurok, TraschenTurokBrandenberger}). Here we extend the USM model to include the contribution of cosmic string loops and implement this extended framework in the CMBACT code. This is done with the objective of quantifying cosmic string loops' contribution to the CMB anisotropies and of improving the accuracy of the computations of the CMB signatures of cosmic string networks. 

This paper is organized as follows. In Sec.~\ref{sec:set}, we compute the stress-energy tensor of a circular cosmic string loop. In Sec.~\ref{sec:model}, we develop the framework necessary to compute the CMB anisotropies generated by cosmic string loops. We start by reviewing the main aspects of the cosmological evolution of a cosmic string network and loop production in Sec.~\ref{sec:vos}. We then extend the USM to also account for the contribution of the cosmic string loops that are produced  throughout the evolution of a cosmic string network in Sec.~\ref{sec:USM}. We characterize the CMB signatures generated by cosmic string loops in Sec.~\ref{sec:loopcont}, study their impact on the anisotropies generated by cosmic string networks in Sec.~\ref{sec:fullspec} and investigate the impact of a reduced intercommutation probability in Sec.~\ref{sec:super}. We then conclude in Sec.~\ref{sec:conc}.

\section{Stress-energy tensor for circular loops\label{sec:set}}

In most situations of interest in cosmology, a cosmic string may be treated as an infinitely thin object that sweeps $1+1$-dimensional worldsheet in spacetime. This worldsheet may be represented by the 4-vector
\be
X^\mu =X^\mu \left(\sigma ^0,\sigma ^1 \right)\,,
\ee
where $\sigma^0$ and $\sigma^1$ are variables that parameterize the string worldsheet. 
In this case, cosmic string dynamics is described by the Nambu-Goto action
\begin{equation}
\label{Action}
S = - \mu_0 \int \sqrt{-\gamma} d^2 \sigma\,,
\end{equation}
where $\mu_0$ is the cosmic string tension --- which, for Nambu-Goto strings, coincides with the mass per unit length --- and is related to the energy-scale of the string forming phase transition. Here $\gamma$ is the determinant of the worldsheet metric $\gamma_{ab}=g_{\mu\nu}X^\mu_{,a} X^\nu_{,b}$ (with $a,b=0,1$), and $g_{\mu\nu}$ is the background metric.

In a Friedmann-Lemaitre-Robertson-Walker (FLRW) background --- with line element 
\begin{equation}
\label{metric}
ds^2 = a(\tau)^2 \left( d \tau^2 - d \textbf{x}^2 \right)\,,
\end{equation}
where $a(\tau)$ is the cosmological scale factor and $d\tau=dt/a$ is the conformal time and $t$ is the physical time --- it is convenient to chose the temporal-transverse gauge:
\be
\sigma^0=\tau,\qquad \mbox{and}\qquad \dot{\mathbf{X}}\cdot\mathbf{X}'=0\,,
\ee
where $X^\mu=(\tau,\mathbf{X})$ and a dot or a prime denotes a derivative with respect to $\sigma^0$ or $\sigma^1$ respectively. In this case, the stress-energy tensor (obtained by varying the action in Eq.~(\ref{Action}) with respect to $g_{\mu \nu}$) may be expressed as
\begin{equation}
\label{SEt}
T^{\mu \nu} = \frac{\mu_0}{\sqrt{-g}} \int   \left( \epsilon \dot{X}^{\mu} \dot{X}^{\nu} - \epsilon^{-1} X^{\prime \, \mu} X^{\prime \, \nu} \right) \delta^{(4)} d^2 \sigma\,,
\end{equation}
where $\delta^{(4)} = \delta^{(4)}(x^{\eta}-X^{\eta}(\sigma^0, \sigma^1))$ 
is a Dirac delta function and $\epsilon^2=\mathbf{X}'^2/(1-\dot{\mathbf{X}}^2)$.

Let us now consider the case of a circular (planar) cosmic string loop with conformal radius $R_c$. For simplicity, we shall assume that the loop has, instantaneously, no radial velocity ($\dot{R}_c \approx 0$). Although cosmic string loops are quite generally expected to oscillate under the effect of their tension, we have verified numerically that this assumption does not have a significant impact on the final results\footnote{See Appendix \ref{ApA} for the stress-energy tensor of a loop with $\dot{R_c}\neq 0$.}. Moreover, we will also assume that the loop has a translational velocity, $v_l$, orthogonal to the loop's plane. In this case, we have that
\begin{equation}
\label{LoopVector}
\textbf{X} = \textbf{X}_i + R_c  \textbf{x} \cos \sigma + R_c  \textbf{y} \sin \sigma + v_l \tau\textbf{z}\,,
\end{equation}
where $\sigma=\sigma^1 \in [0,2 \pi]$, and $\textbf{X}_i$ is the initial location of the center-of-mass of the loop. 
Here, $ \textbf{x}$, $ \textbf{y}$ and  $\textbf{z}$ are three orthogonal unitary vectors defined as
\begin{eqnarray}
\label{UnitVectors}
     & \textbf{x} = \begin{pmatrix}   \sin \theta \sin \phi \\  - \sin \theta \cos \phi \\ \cos \theta  \end{pmatrix}, \\
     & \textbf{y} = \begin{pmatrix}  \cos \phi \cos \psi - \sin \psi \sin \phi \cos \theta \\  \sin \phi \cos \psi + \sin \psi \cos \phi \cos \theta \\ \sin \theta \sin \psi \end{pmatrix} , \\
     & \textbf{z} = \begin{pmatrix}  - \cos \phi \sin \psi - \cos \psi \sin \phi \cos \theta \\  - \sin \phi \sin \psi + \cos \psi \cos \phi \cos \theta \\ \sin \theta \cos \psi  \end{pmatrix} ,
\end{eqnarray}
with $0 \leq \theta < \pi$ and 
$0 \leq \phi, \, \psi < 2 \pi$.

The Fourier transform of the stress-energy tensor of the cosmic string loop is then given by
\begin{equation}
\begin{gathered}
\label{StressEnergFour}
\Theta^{\mu \nu}  =  \mu  \int_{0}^{2 \pi} \text{e}^{i \textbf{k} \cdot \textbf{X}} \left( \epsilon \dot{X}^{\mu} \dot{X}^{\nu} - \epsilon^{-1} X^{\prime \, \mu} X^{\prime \, \nu} \right) d \sigma .
\end{gathered}
\end{equation}
 Assuming, without loss of generality, that $\textbf{k} = k \textbf{k}_{z}$, with $\textbf{k}_{z} = \left\lbrace 0, \; 0, \; 1 \right\rbrace$, we have that
\begin{equation}
\begin{gathered}
\label{ScalarProduct}
\textbf{X} \cdot \textbf{k} = k \textbf{X}_i \cdot \textbf{k}_z + v_l \tau z_{z} + R k A \sin (\sigma+B)\,,
\end{gathered}
\end{equation}
where 
\begin{equation}
\label{ABpar}
A^2 = x_{z}^2+y_{z}^2 \,,\qquad \tan B = \frac{x_{z}}{y_{z}}\,,
\end{equation}
and the subscript `z' denotes the projection along the $\textbf{k}_{z}$ direction. The real part of ``$00$''-component of the stress-energy tensor (\ref{StressEnergFour}) may then be written as
\begin{equation}
\begin{gathered}
\label{Theta00}
\Theta^{0 0} = M  J_0 (\mathcal{X}) \cos \varphi_0,
\end{gathered}
\end{equation}
where $M = 2 \pi \mu_0 R_c \gamma_v $, 
$\varphi_0 = k  \textbf{X}_i \cdot \textbf{k}_z + v_l k \tau z_{z} $, $\mathcal{X} = k R_c A $,
$\gamma_v = (1-v_l^2)^{-1/2}$, and $J_n(..)$ is
a Bessel function of the first kind.

The spatial components of the stress-energy tensor are given by
\begin{equation}
\begin{gathered}
\label{Thetaij}
\Theta^{i j} = \Theta^{0 0} \times \\
 \left[ v_l^2 z^i z^j -   \frac{\gamma_v^{-2}}{2} \left( x^i x^j \mathcal{I}_- + y^i y^j  \mathcal{I}_+ + 2 \mathcal{I} x^{(i} y^{j)} \right) \right],
\end{gathered}
\end{equation}
where 
\begin{equation}
\begin{gathered}
\label{Int}
\mathcal{I}_{\pm} = 1 \pm \frac{J_2(\mathcal{X})}{J_0(\mathcal{X})} \cos 2 B, \qquad \mathcal{I} =  \frac{J_2(\mathcal{X})}{J_0(\mathcal{X})} \sin 2 B, \\
\mbox{and} \qquad y^{(i} x^{j)} = \frac{1}{2} \left( y^{i} x^{j} + x^{i} y^{j} \right).
\end{gathered}
\end{equation}
The scalar, vector and tensor components of the stress-energy 
tensor (\ref{Thetaij}) are given, respectively, by 

\begin{equation}
\begin{gathered}
\label{SVT}
\Theta^S = \left( 2 \Theta^{33} - \Theta^{11} - \Theta^{22} \right)/2,\\
\Theta^V = \Theta^{13},\\
\Theta^T = \Theta^{12},
\end{gathered}
\end{equation}
while the trace $\Theta = \Theta_{ii}$ and velocity field $\Theta^D = \Theta_{03}$ are fixed by imposing local 
energy-momentum conservation~\cite{AlbrechtBattyeRobinson}.

\section{Modeling the cosmic string networks with loops\label{sec:model}}

A cosmic string network has two primary constituents: long strings --- cosmic strings that stretch beyond the horizon --- and subhorizon closed string loops. The creation of loops happens persistently throughout the evolution of a cosmic string network due to string collisions. These loops detach from the long string network and evolve independently from it. Thus, there is a continual energy loss by the long string network that plays a crucial role in its dynamics. In this section, we review the main aspects of cosmic string network dynamics and loop production and extend the USM model to account for cosmic string loops.

\subsection{Cosmic string network evolution and loop production\label{sec:vos}}

The evolution of topological defect networks has been extensively studied using numerical 
\cite{MooreShellardMartins, MartinsShellard2006, BlancoPilladoOlumShlaer, HindmarshLizarragaUrrestillaDaverioKunz, CorreiaMartins2} and semi-analytical~\cite{MartinsShellard, MartinsShellard2,AvelinoMartins,Avelino:2011ev,Sousa:2011ew,SousaAvelino2, SousaAvelino3} methods. The Velocity-dependent One-Scale (VOS) model, in particular, provides a simple and yet informative description of the large-scale dynamics of defect networks that grasps the main features of averaged network evolution. This model --- initially introduced for cosmic strings~\cite{MartinsShellard, MartinsShellard2} but later generalized to  defects of arbitrary dimensionality~\cite{AvelinoMartins,Avelino:2011ev,Sousa:2011ew,SousaAvelino2, SousaAvelino3} --- provides a quantitative description of the evolution of a long string network by its root-mean-squared velocity (RMS) $\vv$ and characteristic conformal length $L_c$~\cite{MartinsShellard}:
\bq
\frac{d \vv}{d \tau} & = & (1-\vv^2) \left[ \frac{k(\vv)}{L_c} - 2 \frac{\dot{a}}{a} \vv \right]\,,\\
\frac{d L_c}{d \tau} & = &  \frac{\dot{a}}{a} L_c \vv^2 + \frac{\cc}{2} \vv\,,\label{VOSL}
\eq
where $\cc$ is a parameter that quantifies the loop-chopping efficiency, and $k(\vv)$ is a momentum paremeter. Nambu-Goto simulations are well described by $\cc=0.23$ and a momentum parameter of the form~\cite{MartinsShellard2}
\begin{equation}
\label{MomentPar}
k(\vv) = \frac{2\sqrt{2}}{\pi} \frac{1-8 \vv^6}{1+\vv^6} (1-\vv^2) (1+ 2 \sqrt{2} \vv^3),
\end{equation}
see Ref.~\cite{CorreiaMartins} for the latest calibration of these parameters from the simulation of Abelian-Higgs cosmic strings.

Since the main objective of the present work is to study the potential impact of cosmic string loops on the CMB, we also need to describe the number (and length) of loops  produced throughout cosmic history. This subject has had considerable attention in the literature (see e.g.~\cite{Kibble:1984hp,Caldwell:1991jj,DePies:2007bm,Sanidas:2012ee,Kuroyanagi:2012wm,Sousa:2013aaa,Blanco-Pillado:2013qja,Sousa:2014gka,Sousa:2020sxs}) since cosmic string loops are expected to give rise to a stochastic gravitational wave background that is expected to be within reach of gravitational wave experiments in the near future. Here we shall adopt the semi-analytical approach of Ref.~\cite{Sousa:2013aaa} since it allows for the characterization of the number of loops created throughout the realistic cosmological history (even through the radiation-matter and matter-dark-energy transitions). In this approach, it is implicitly assumed that the main energy-loss mechanism in a cosmic string network is the creation of loops. Thus, all the energy lost by the network (besides the loss that results from Hubble expansion) goes into the formation of loops. The characteristic length of the network $L_c$ may be regarded as a measure of the energy density of the network
\begin{equation}
\label{DensEnergL}
\rho = \frac{\mu_0}{a^2 L_c^2}\,,
\end{equation}
where $\rho$ is the average energy density of the network. Thus, using Eq.~(\ref{VOSL}), one finds that the energy density lost by the network is given by
\be
\label{energyloss}
\left.\frac{d\rho}{dt}\right|_{\rm{loops}}=\cc\frac{\vv}{aL_c}\rho\,.
\ee

In this approach, it is also generally assumed that cosmic string loops are created with a length that is a fixed fraction of the characteristic length of the network at the time of creation
\begin{equation}
\label{LoopSize}
l_c^b = \alpha L_c(t_b)\,,
\end{equation}
where $0<\alpha<1$ is a constant loop-size parameter and $l_c^b$ is the comoving length of the cosmic string loop at its time of birth $t_b$. The loop-size parameter $\alpha$ may be calibrated using numerical simulations. Note, however, that numerical simulations are not conclusive as to the length of the loops produced in a cosmic string network's evolution. Nambu-Goto simulations consistently indicate that about $10\%$ of the energy lost by the long string network goes into the formation of large loops with $\alpha\sim 0.34$~\cite{Blanco-Pillado:2013qja,Lorenz:2010sm,Blanco-Pillado:2019tbi}\footnote{There is, however, a severe disagreement in the number of small loops predicted by simulation-inferred models developed by different groups~\cite{Auclair:2019wcv} (see, however,~\cite{Blanco-Pillado:2019vcs,Blanco-Pillado:2019tbi}).}. Meanwhile, Abelian-Higgs simulations suggest much smaller density of loops due to an additional mechanism of energy loss: the emission of scalar and gauge radiation~\cite{HindmarshStuckeyBevis, HindmarshLizarragaUrrestillaDaverioKunz}. Here, we shall treat $\alpha$ as a free parameter of the model in order to study different scenarios. The number density of loops created per unit time is then given by~\cite{Sousa:2013aaa}:
\be
\frac{dn_l}{dt}=\frac{1}{a l_c^b}\left.\frac{d\rho}{dt}\right|_{\rm{loops}}=\frac{\cc}{\alpha}\frac{\vv}{a^4 L_c^4}\,.
\label{loopprod}
\ee
Although, in reality, one does not expect all loops to be created with exactly the same length, the effect of having a distribution of lengths at the moment of creation may, to some extent, be included in this model through a renormalization of Eq.~(\ref{loopprod}) by a factor $\mathcal{F}$~\cite{Sanidas:2012ee,Blanco-Pillado:2013qja,Sousa:2020sxs}.

After creation, cosmic string loops are expected to emit gravitational radiation at a roughly constant rate
\begin{equation}
\label{LoopEnerLose}
\left.\frac{d E_l}{d t}\right|_{\rm{gr}} = - \Gamma G \mu_0^2,
\end{equation}
where $\Gamma \sim 50$ 
\cite{QuashnockSpergel, ScherrerQuashnockSpergel},
\begin{equation}
\label{Energy}
E_l = \mu_0 a \int \epsilon d \sigma = \mu_0 l
\end{equation}
is the energy of the loop, and $l = a l_c$ is the length of the cosmic string loop. Loops then shrink as a result of this emission until they eventually evaporate. In fact, using Eqs.~(\ref{LoopSize})-(\ref{Energy}), we find that the comoving radius of the cosmic string loop evolves as
\begin{equation}
\label{Radius}
R_c(\tau) = \frac{ \alpha L_c(\tau_i) a(\tau_i) - \Gamma  G \mu_0 \left( t(\tau)-t(\tau_i) \right) }{{2 \pi a(\tau)}}\,,
\end{equation}
where $\tau_i$ is the conformal time of birth of the loops and $\tau_f$ is the time of loop decay (for which $R_c(\tau_f) = 0$).

\subsection{An Unconnected Segment Model with Loops\label{sec:USM}}

This section extends the  USM~\cite{AlbrechtBattyeRobinson,PogosianVachaspati} --- which describes the stress-energy tensor of a cosmic string network --- to also account for cosmic string loops. 
In the USM, the long string network is modeled as a collection of uncorrelated, straight finite segments created simultaneously at an early time. The positions of the segments are drawn from a uniform distribution in space, and the direction of their velocities --- which is assumed to be orthogonal to the string itself --- is chosen from a uniform distribution on a two-sphere. The VOS model is then used to set the comoving length of the segments $L_c$ and the magnitude of their velocity $\vv$. Since our model also includes the contribution of long strings (besides that of loops), we preserve the main features of this model.

To account for the energy loss caused by loops production, a fraction of the segments decays at each time instant $\tau_i$:
\be
N(\tau_i)=\mathcal{V}\left[n(\tau_{i-1})-n(\tau_i)\right]\,,
\ee
where $N(\tau_i)$ is the number of long string segments that decay between $\tau_{i-1}$ and $\tau_i$, $\mathcal{V}$ is the simulation volume, and $n(\tau)$ is the number density of long strings
\be
n(\tau)=\frac{C(\tau)}{L_c(\tau)^3}\,. 
\ee
Agreement between the number density of strings in this model and that predicted by the VOS model is ensured by requiring that the normalization function $C(\tau)$ is given by $\mathcal{V}/L_c(\tau)^3$ at any given conformal time $\tau$.

In order to include loops in this model, we assume that the segments that decay at a given time are ``converted'' into cosmic string loops and, thus, the number of loops created at a given time is given by
\be
N_l(\tau_i)=\frac{\mu_0 N(\tau_i) L_c(\tau_i)}{\mu_0 l_c^b(\tau_i)}=\frac{N(\tau_i)}{\alpha}\,,
\ee
where we have used Eq.~(\ref{LoopSize}). This ensures that there is a balance between the energy lost by the cosmic string network and the total energy of the loops created, so that the number of loops created is in agreement with Eq.~(\ref{loopprod}).

In this extension of the USM, the long string segments decay at each (discrete) time instant $\tau_i$ and a population of $N_l(\tau_i)$ circular loops is created with an initial comoving radius $R_c(\tau_i)$ given by 
Eq.~(\ref{Radius}) and a stress-energy tensor given by Eqs.~(\ref{Theta00})-(\ref{Thetaij}). These loops then shrink (by emitting gravitational waves) according to Eq.~(\ref{Radius}) until they eventually disappear at a time $\tau_f$ (in which $R_c(\tau_f)=0$). The appearance/disappearance of the cosmic string loops is achieved in the same manner as the decay of string segments in the original USM model. In fact, the total stress-energy tensor of the loop network is written as
\begin{equation}
\label{AllLoops}
\tilde{\Theta}^{\mu \nu}_{
\text{L}}(\textbf{k}, \tau) = \sum_{n}^{N_T} \Theta^{\mu \nu}_n (\textbf{k}, \tau) T^{\text{off}}(\tau,\tau_f^n) T^{\text{on}}(\tau,\tau_i^n)\,,
\end{equation}
where $\Theta^{\mu \nu}_n (\textbf{k}, \tau)$ is the stress-energy tensor of the $n$-th loop, $N_T$ is the total number of loops, and $\tau_i^n$ and $\tau_f^n$ are respectively the conformal times of creation and evaporation of the $n$-th loop. Here, we have also introduced the functions
\begin{equation}
\label{Toff}
T^{\text{off}} (\tau,\tau_f)=   
\begin{cases}
    1  &  \tau < \lambda_- \tau_f \\
    \frac{1}{2} + \frac{1}{4} \left(x_{\text{off}}^3 - 3 x_{\text{off}} \right)  & \lambda_- \tau_f \le \tau < \tau_f \,,\\
    0 & \tau_f\le\tau
\end{cases}
\end{equation}
where 
\begin{equation}
\label{xoff}
x_{\text{off}} = 2 \frac{ \ln(\lambda_- \tau_f / \tau) }{\ln (\lambda_-)} - 1\,,
\end{equation}
and
\begin{equation}
\label{Ton}
T^{\text{on}} (\tau,\tau_i)=   
\begin{cases}
    0  &  \tau < \tau_i \\
    \frac{1}{2} + \frac{1}{4} \left(3 x_{\text{on}} - x_{\text{on}}^3 \right)  &  \tau_i \le \tau < \lambda_+ \tau_i\,, \\
    1 & \lambda_+ \tau_i\le\tau
\end{cases}
\end{equation}
where 
\begin{equation}
\label{yon}
x_{\text{on}} = 2 \frac{\ln( \tau_i / \tau )}{\ln(1/\lambda_+)}-1\,.
\end{equation}
The $T^{\text{off}}$ function is responsible for ``turning off'' the contribution to the stress-energy tensor of loops that have already evaporated at the time $\tau_f$ and it is identical to the $T^{\text{off}}$ function used in the original USM to model the decay of long string segments. We have also included the $T^{\text{on}}$ function to ``turn on'' the contributions of all the loops created at the time $\tau_i$\footnote{The USM, in its original implementation~\cite{AlbrechtBattyeRobinson}, actually included, for computational efficiency, a slightly different $T^{\text{on}}$ function to only ``turn on'' segments once they may contribute significantly to the CMB anisotropies. This was, however, abandoned in later extensions of the model.}. These functions then ensure that $\tilde{\Theta}^{\mu \nu}_{\text{Loops}}(\textbf{k}, \tau)$ only has a contribution from the relevant loop populations: those that were created at a time $\tau_i<\tau$ and have not evaporated completely yet at time $\tau$. The constants $\lambda_\pm$ determine how fast loops appear or disappear: in fact, $T^{\text{on}}$ ($T^{\text{off}}$) grows (decreases) continuously from 0 (1) to 1 (0) between $\tau_i$ ($\lambda_- \tau_f$) and $\lambda_+ \tau_i$ ($\tau_f$). Here, we take $\lambda_\pm=1\pm0.2$\footnote{We have verified numerically that the values of $\lambda_\pm$ do not have a significant impact on the final results}.

An essential feature of the USM for long strings is that, to ensure computational efficiency, all cosmic string segments that decay at a given time are \textit{consolidated} into a single string segment. In fact, since the segments are distributed randomly in real space, their random positions correspond to a random phase in Fourier space. Thus, the amplitude of the sum of their contribution to the stress-energy tensor is essentially a 2-dimensional random walk. As a result, the total stress-energy tensor in Fourier space is simply the stress-energy tensor of a single segment weighted by a factor of $\sqrt{N}$
\cite{AlbrechtBattyeRobinson}. Naturally, for numerical efficiency, we shall preserve this feature and consolidate each loop population --- i.e. the loops that are created (and decay) at the same time --- into a single cosmic string loop. However, cosmic string loops cannot realistically be expected to follow a random distribution in space. As a matter of fact, as discussed in Ref.~\cite{Wu:1998mr}, the positions of loops are highly correlated with the positions of long strings: loops are created along the strings, and they tend to move in the same direction as the string from which they are chopped. Assuming that loops are randomly distributed in space and move in random directions is, therefore, not realistic and may have a significant impact on the results. However, in any case, we can consolidate a loop population into a single loop if we do so in two steps. At any given time $\tau_i$, $N(\tau_i)$ segments decay into loops, and thus, we have, on average, $\alpha^{-1}$ loops created per decaying string segment. Given the strong correlation between the positions of strings and loops, we may then expect that these $\alpha^{-1}$ loops are created at random positions along the decaying string and thus the sum of their contributions to the stress-energy tensor in Fourier space corresponds to a sum of terms with random phases. Similarly to long strings, these may then be consolidated into a single loop located at the position of the center-of-mass of the decaying string at the time of creation $\tau_i$, with a weight $1/\sqrt{\alpha}$. After this first consolidation step, we have $N(\tau_i)$ loops that were created at the same time $\tau_i$ and located at the (random) positions of the centers-of-mass of the decaying segments. We may then consolidate these loops into a unique loop with a weight of $\sqrt{N/\alpha}$:
\be
\begin{gathered}
\label{AllLoops2}
\tilde{\Theta}^{\mu \nu}_{
\text{L}} (\textbf{k}, \tau) = \\
= \sum_j \sqrt{N_l(\tau_i^j)}\Theta^{\mu\nu}_j(\kk,\tau) T^{\rm off}(\tau,\tau_f^j) T^{\rm on}(\tau,\tau_i^j)\,,
\end{gathered}
\ee
where the index $j$ runs over the consolidated loops.

The correlation between loops and long strings is taken into account by positioning the  consolidated loop --- that represents the population of loops created at $\tau_i$ --- at the position of the consolidated decaying string segment at the time of decay and by imposing that the direction of the velocity of the loop coincides with that of the decaying string. We should then have
\begin{equation}
\label{LocLoop}
\varphi_0 = k  \textbf{X}_0 \cdot \textbf{k}_z +  k \tau_i z_{z} \vv +  v_l k (\tau - \tau_i) z_{z}\,,
\end{equation}
where $ \textbf{X}_0$ is the (randomly assigned) initial position of the decaying consolidated segment, $\textbf{X}_0 \cdot \textbf{k}_z +  k \tau_i z_{z} \vv$ is its position at the time of decay. Since the center-of-mass velocity of the loop is expected to scale as $\gamma_v v_l \propto a^{-1}$ due to the expansion of the background, the velocity of the loop is given by
\begin{equation}
\label{LoopVel}
v_{l}(\tau) = \frac{v_{l}(\tau_i)}{\sqrt{v_{l}^2(\tau_i)+ \left( 1-v^2_{l}(\tau_i) \right) \left(\frac{a(\tau)}{a(\tau_i)}\right)^2}}  
\end{equation} 
at any instant of time $\tau>\tau_i$.

\section{Cosmic Microwave Background anisotropies\label{sec:anisotropies}}

Although the CMB has a nearly perfect black body spectrum with an approximately uniform temperature, there are tiny temperature fluctuations across the sky~\cite{Fixsen:1996nj}. The CMB is generally characterized in terms of the angular power spectrum, $C_\ell$, of the temperature fluctuations
\be
C_\ell=\frac{1}{2\ell+1}\sum_{m=-1}^\ell\left<a^*_{\ell m}a_{\ell m} \right>\,,
\ee
where angled brackets represent an ensemble average. Here, $a_{\ell m}$ are the coefficients of the decomposition of the temperature fluctuations, $\triangle(\mathbf{\hat{n}})=\triangle T/T$, into spherical harmonics
\be
\triangle(\mathbf{\hat{n}})=\sum_ {\ell m}a_{\ell m}Y_{\ell m}(\mathbf{\hat{n}})\,,
\ee
where $\mathbf{\hat{n}}$ is the direction of the line of sight and $Y_{\ell m}$ are spherical harmonic functions. The angular power spectrum, then, allows us to separate the contributions to different angular scales of the CMB anisotropies.

Here, we are also going to compute the Cold Dark Matter (CDM) linear power spectrum,
\be 
P(k)=\left|\delta^2\left(\mathbf{k}\right)\right|\,,
\ee
where $\delta\left(\mathbf{k}\right)$ is the Fourier transform of the density contrast,
\be 
\delta\left(\mathbf{x}\right)=\frac{\rho_m(\mathbf{x})-\left<\rho_m\right>}{\left<\rho_m\right>}\,,
\ee
where $\rho_m(\mathbf{x})$ is the matter density at a given position $\mathbf{x}$ and $\left<\rho_m\right>$ is its average value.

In this section, we compute the CMB and linear CDM power spectra generated by cosmic string networks with loops. To do this, we extend the publicly available CMBACT code to also account for cosmic string loops, by implementing the modifications described in the previous sections. Our results are obtained by averaging over 500 realizations of a brownian  cosmic string network with the following cosmological
parameters: $\Omega^0_b h^2=0.0224$, $\Omega^0_m h^2=0.1424$ for baryon and matter density parameters, and $H_0 = 100 h \, \text{kms}^{-1} \text{Mpc}^{-1}$, with $h=0.674$ 
for the Hubble parameter at the present time~\cite{Planck2020}. The tension of cosmic strings is 
fixed to $G \mu_0 =10^{-7}$ and we assume that all loops are created with the same length (and thus $\mathcal{F}=1$) unless stated otherwise. The unmodified CMBACT is used to obtain the power spectra generated by long strings only.

\subsection{The contribution of cosmic string loops\label{sec:loopcont}}

Before going into the CMB anisotropies generated by the full cosmic string network, with both long strings and loops, we start by presenting the contribution that comes solely from cosmic string loops. To do so, we modify the CMBACT code in such a way as to include only the contribution of cosmic string loops to the stress-energy tensor (\ref{AllLoops2}).

\begin{figure} [h!]
\begin{center}
\includegraphics[width=3.4in]{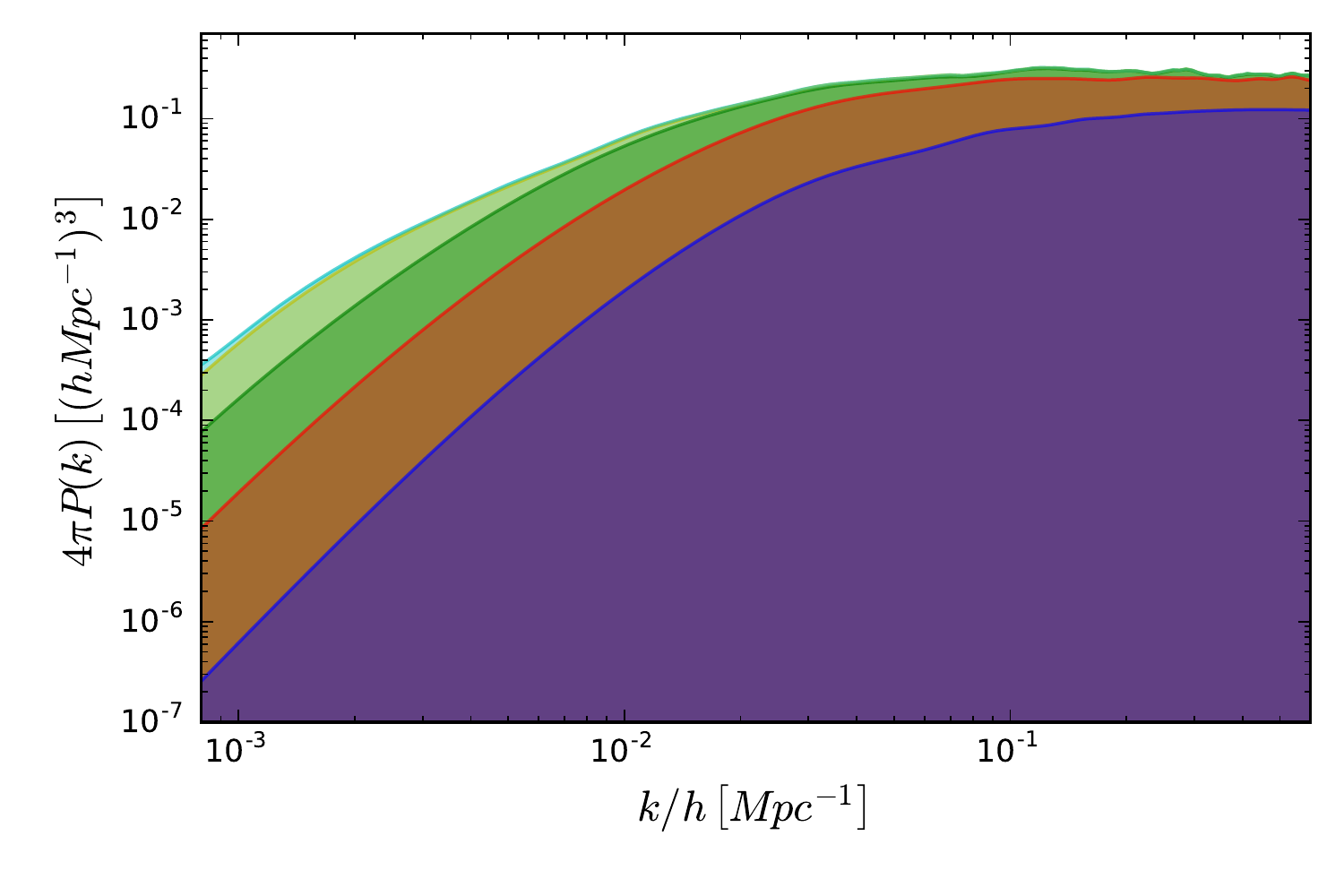}
\caption{Linear CDM power spectrum generated by cosmic string loops. We include the CDM power spectra generated by cosmic string loops up to $a=10^{-4}$ (blue line),$a=10^{-3}$ (red line), $a=10^{-2}$ (green line), $a=10^{-1}$ (olive line), and $a=1$ (cyan line).  We have averaged over 500 realizations of cosmic string loop networks, and took $G\mu_0=10^{-7}$ and $\alpha=10^{-1}$.\label{pkevo}}
\end{center}
\end{figure}

Let us start by looking into the linear CDM power spectrum. To have a clear picture of the contribution of cosmic string loops, we have studied the evolution of the linear CDM power spectrum generated by loops. In particular, in Fig.~\ref{pkevo}, we plot the linear CDM power spectrum generated by cosmic string loops up until different epochs in cosmic history (namely until a scale factor of $a = 10^{-4}$, $a = 10^{-3}$, $a = 10^{-2}$, $a = 10^{-1}$, and $a = 1$, corresponding to the present time). Therein, we may see that, as time progresses, cosmic string loops contribute dominantly at increasingly larger scales (smaller values of $k$). As a matter of fact, since the correlation length of the cosmic string network increases throughout the evolution, the radius of the loops produced also increases. However, the number of loops created throughout the evolution decreases with time because the network becomes progressively less dense. As a result, the dominant contribution to the CDM power spectrum comes from loops created in the radiation era or around the radiation-matter transition.

\begin{figure} [h!]
\begin{center}
\includegraphics[width=3.4in]{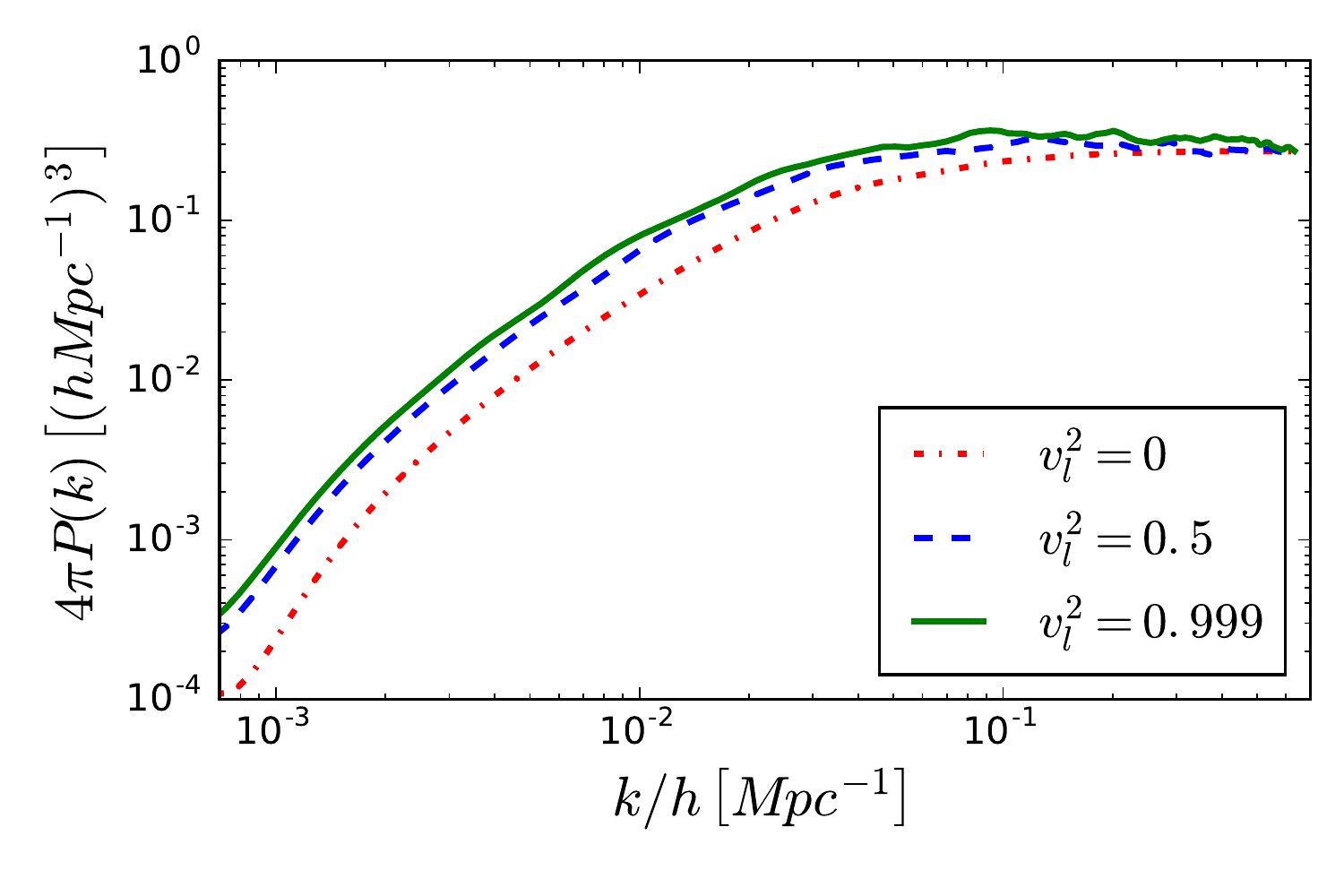}
\caption{\label{pkvel} Linear CDM power spectrum generated by cosmic string loops with different velocities. We chose $G\mu_0=10^{-7}$ and $\alpha=10^{-1}$ and averaged over 500 realizations of cosmic string and/or loop network realizations.}
\end{center}
\end{figure}

\begin{figure} [h!]
\begin{center}
\includegraphics[width=3.4in]{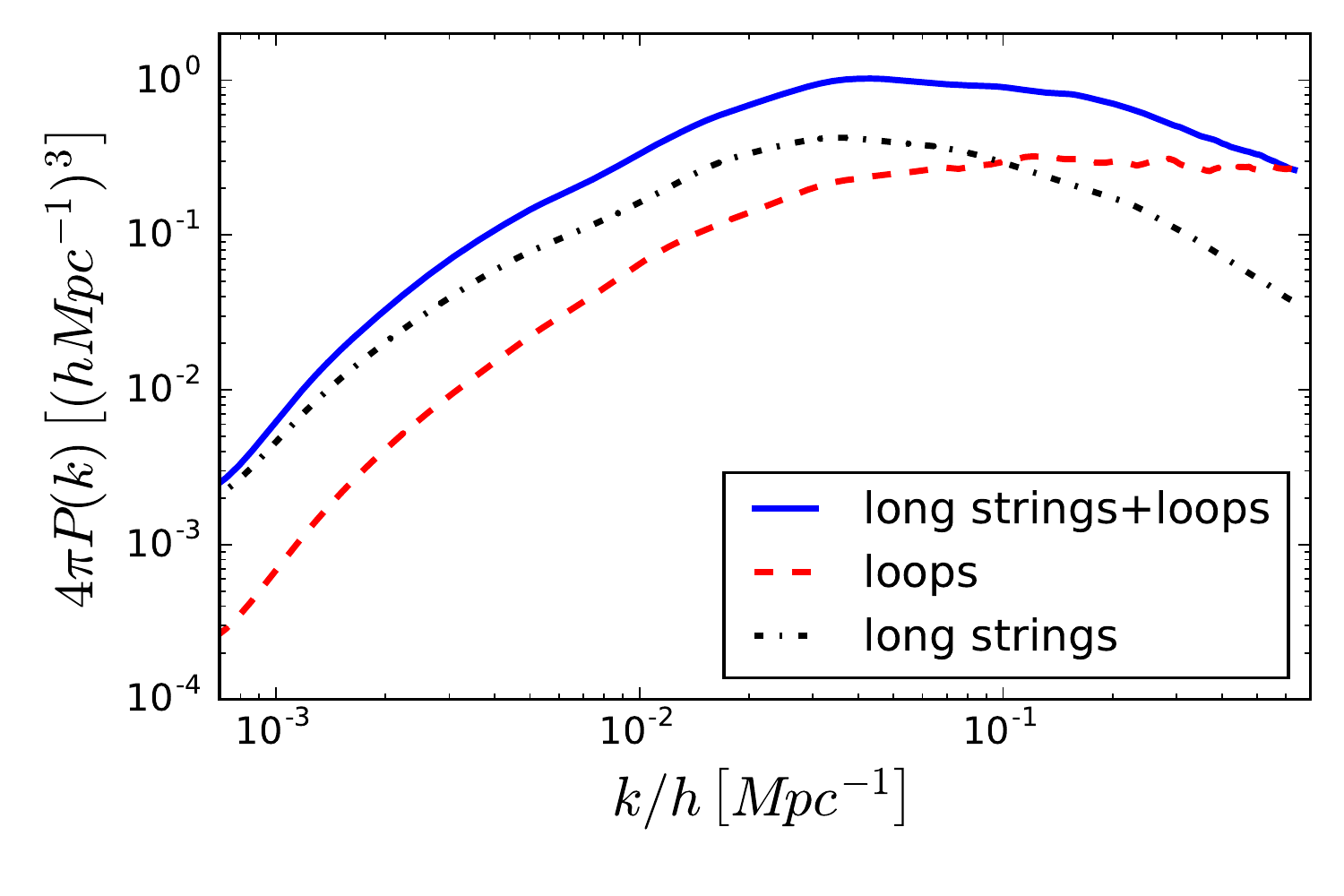}
\caption{\label{pkcor} Linear CDM power spectrum generated by a cosmic string network with loops. The solid (blue) line represents the power spectrum generated by cosmic strings and cosmic string loops, while the dash-dotted (black) and dashed (red) lines represent the contributions of long strings and cosmic string loops. We chose $G\mu_0=10^{-7}$ and $\alpha=10^{-1}$ and averaged over 500 realizations of cosmic string and/or loop network realizations.}
\end{center}
\end{figure}

\begin{figure}[h!]
\begin{center}
\includegraphics[width=3.5in]{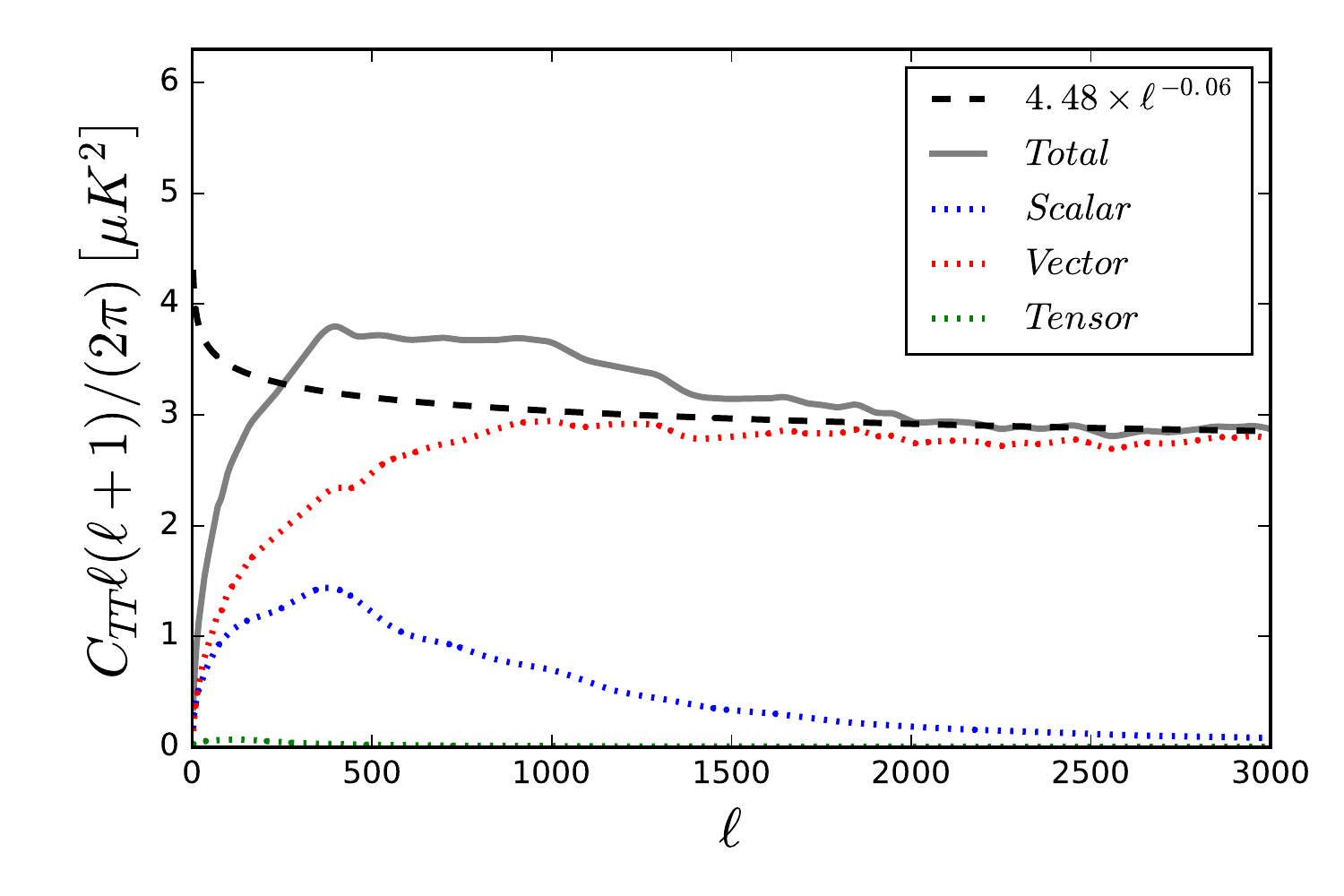}
\caption{\label{Fig:TT} Temperature angular power spectrum generated by cosmic string loops and its scalar, vector, tensor components. For $\ell>1500$ the total contribution is well approximated by $C_{TT} \frac{\ell(\ell+1)}{2 \pi} \sim \ell^{-0.05}$, i.e it is almost constant due to the vector contribution. }
\end{center}
\end{figure}


In Fig.~\ref{pkvel}, we plot the linear power spectrum generated by randomly distributed cosmic string loops for different values of the loop translational velocity. Therein, one may see that for static loops the power spectrum is flat for sufficiently large $k$ (or small scales) as predicted in~\cite{Wu:1998mr}. This is mainly a result of the fact that static loops act, on sufficiently large scales, effectively as point-like sources of perturbations. On the other hand, if the loops are moving, they generate filament-like perturbations, which causes a transfer of power towards larger scales (smaller $k$). Although this effect is more accentuated for larger value of $v_l$, our results show that the amplitude and shape of the linear power spectrum does not depend strongly on the magnitude of the velocity of loops. However, this figure also shows that, for non-vanishing $v_l$, the spectrum starts to develop a decreasing slope at small scales, exhibiting signs of the transition to the $k^{-1}$ regime predicted in Ref.~\cite{Wu:1998mr} for even larger values of $k$.

As Fig.~\ref{pkcor}, where we plot the linear CDM power spectrum of loops alongside that of long strings, shows, except for the slower decrease at small scales, the spectrum generated by cosmic string loops is very similar both in amplitude and in shape to that generated by long strings. Both have peaked power spectra due to the enhancement of perturbations caused by the fact that there is a large number of sources with approximately the same length at roughly the same distance; however, in the case of loops,
the peak is located at larger $k$ because the length of the loops is a fraction of the correlation length of long strings. Moreover, since loops decay after formation, there is some dispersion in the length of the sources of perturbation, and, as a result, the peak of the spectrum is broader and less pronounced.

The differences in the CDM linear power spectrum of long strings and loops naturally translate into differences in the CMB anisotropies. In Fig.~\ref{cmbevo} the TT, EE, TE and BB components of the CMB angular power spectra generated by cosmic string loops are plotted up to different cosmological scale factors. The shape of angular spectra for cosmic loops is also very similar to that generated by long string networks (cf. Fig.~\ref{cmbcor}), but its maximum amplitude is about one order of magnitude smaller, and the peaks of the spectra also appear at a smaller angular scale (or larger multipole $\ell$). The most noticeable difference, however, appears in the vector components. As a matter of fact, the vector contribution to the temperature anisotropies does not decrease with the increase of multipole moment $\ell$ (decreasing angular scales) as it happens for long strings. We anticipate that small circular loops are responsible for this effect: due to their shape, they actively generate rotational movements of matter, giving rise to the divergenceless (vortical) velocity field. In particular, for $\ell>1500$, the vector contribution to the TT angular power spectrum generated by loops is approximately constant, as shown in Fig.~\ref{Fig:TT}, while for long strings $C_{TT} \frac{\ell(\ell+1)}{2 \pi} \sim \ell^{-1.5}$\cite{PogosianTyeWassermanWyman2}.

Fig.~\ref{cmbevo} also shows that cosmic string loops generate temperature anisotropies at progressively larger scales between the epoch of the last scattering and the present time since the length of loops (at the moment of creation) increases with time. However, since the number of created loops decreases roughly as $t^{-4}$, the dominant contribution (corresponding to the peak of the spectrum) is generated earlier in cosmological history. The polarization anisotropies --- as is the case for long strings and domain walls --- are created mainly in two epochs: the dominant peak at small angular scales is generated around the last scattering ($a\sim 10^{-3}$), while the large-scale subdominant peak is created around the epoch of reionization ($a\sim 10^{-1}$).

The CMB power spectra generated by cosmic string loops with different translational velocities ($v_l^2=0$, $v_l^2=0.5$, 
$v_l^2=0.999$) are plotted in Fig.~\ref{cmbvel}. The results show that, although the general shape of the spectrum does not depend 
significantly on the velocity of loops 
(see Fig.\ref{pkvel}), 
the amplitude of the anisotropies does. The temperature anisotropies, in particular, are highly dependent on the speed of loops and generally increase with increasing $v_l$. Vector and tensor modes, due to their nature, are more affected, while 
scalar modes do not have such significant change, as shown in Fig \ref{cmbvel}. 
Note however that, in general, we do not expect loops to be static or non-relativistic. In fact, they are expected to move with relativistic speeds and smaller loops to have higher velocities in general. Here and through the remainder of this paper, we shall take $v_l^2=0.5$ unless stated otherwise. Since the contribution of loops is expected to be subdominant when compared to that of long strings, this assumption is not expected to have a significant impact on the final results.

\subsection{CMB anistropies generated by cosmic string networks with loops\label{sec:fullspec}}

In this section we will characterize the CMB anisotropies generated by a cosmic string network, including both loops and long strings. To do so, we implemented the extension of the USM described in Sec.~\ref{sec:USM} in the CMBACT code. In particular, we write the total stress-energy tensor of the network as a sum of the contributions of loops and long strings,
\begin{equation}
\label{SEtFull}
\Theta_{\text{Network}}^{\mu \nu} (\textbf{k},\tau) = \Theta_{\text{S}}^{\mu \nu} (\textbf{k},\tau) + \tilde{\Theta}_{\text{L}}^{\mu \nu} (\textbf{k},\tau)\,,
\end{equation}
where $\Theta_{\text{S}}^{\mu \nu} (\textbf{k},\tau)$ is the stress-energy tensor of long strings as described in the original USM~\cite{AlbrechtBattyeRobinson,PogosianVachaspati} (which includes the contribution of all consolidated string segments) and $\tilde{\Theta}_{\text{L}}^{\mu \nu} (\textbf{k},\tau)$ is given by Eq.~(\ref{AllLoops}). As discussed in the previous section, numerical simulations show a strong correlation between the positions and velocities of long strings and cosmic string loops~\cite{Wu:1998mr}. This vital ingredient is taken into account in this computation by imposing that the loops are created where a consolidated long string has decayed. Moreover, the loop velocity $v_l$ is orientated along the same direction as the long string velocity $\bar{v}$, as it is encoded in Eq.~(\ref{LocLoop}).

\begin{widetext}

\begin{figure}[h!]
\begin{center}
\includegraphics[width=6.7in]{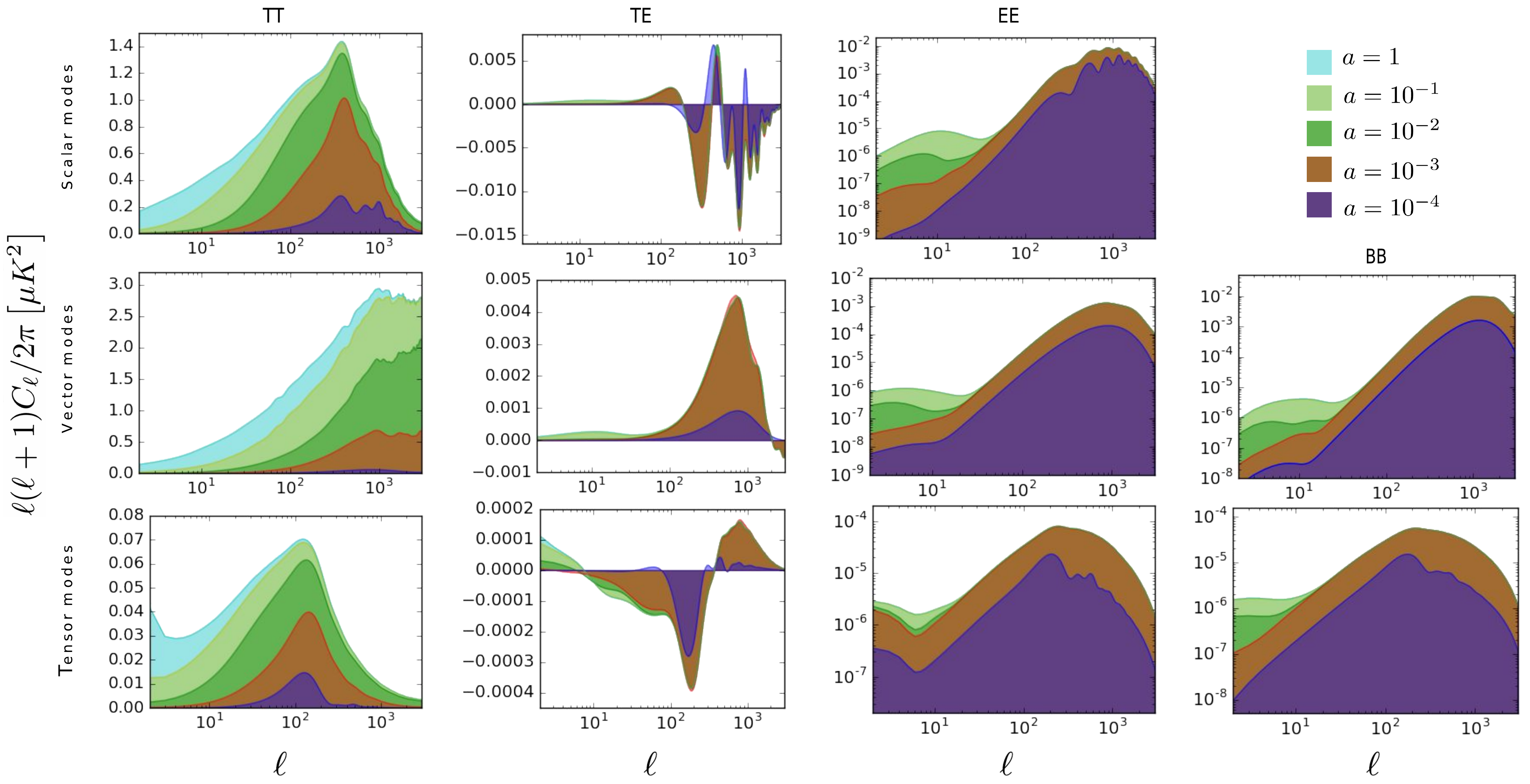}
\caption{CMB anisotropies generated by cosmic string loops. From left to right, we plot  the TT,TE, EE and BB power spectra, as a function of the multipole moment $\ell$. The top, middle and bottom rows represent the scalar,vector and tensor components, respectively. In each of the plots we include the angular power spectra generated by cosmic string loops until $a=10^{-4}$ (blue line), $a=10^{-3}$ (red line), $a=10^{-2}$ (green line), $a=10^{-1}$ (olive line), and $a=1$ (cyan line).  We have averaged over 500 realizations of cosmic string loop networks, and took $G \mu_0=10^{-7}$, $\alpha=10^{-1}$. \label{cmbevo}}
\end{center}
\end{figure}



\begin{figure}[h!]
\begin{center}
\includegraphics[width=6.7in]{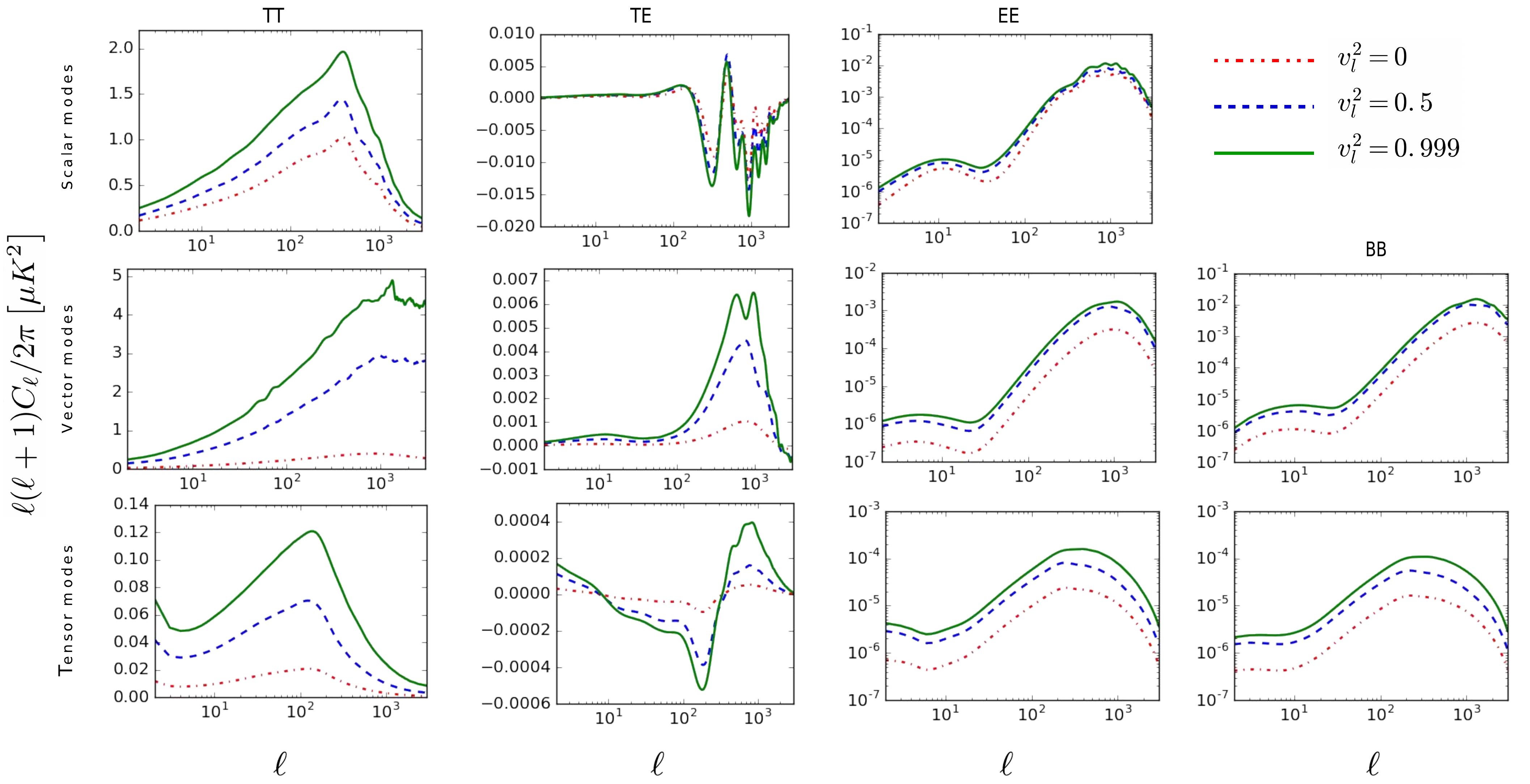}
\caption{CMB anisotropies generated by cosmic string loops with different velocities. From left to right, we plot the TT, TE, EE and BB power spectra, as a function of the multipole moment $\ell$. The top, middle and bottom rows represent the scalar, vector and tensor components, respectively. Each of the plots represents different values of translation velocities $v_{l}^2 =  0$ (dash-dotted red line), $v_{l}^2 =  0.5$ (dashed blue line), $v_{l}^2 =  0.999$ (solid green line), while $G\mu=10^{-7}$, $\alpha=10^{-1}$ and the results are obtained by averaging over 500 realizations.\label{cmbvel}}
\end{center}
\end{figure}



\begin{figure}[h!]
\begin{center}
\includegraphics[width=6.7in]{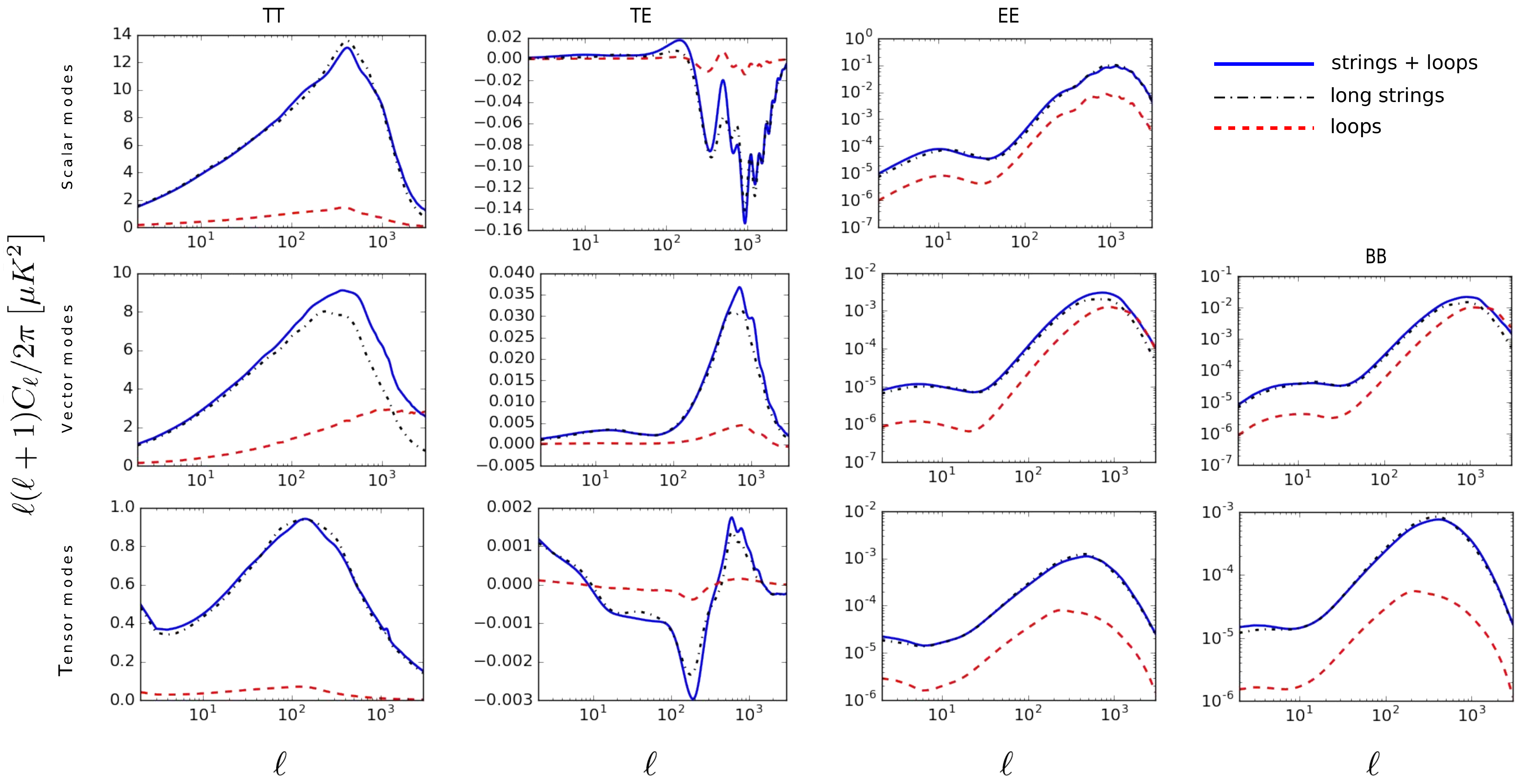}
\caption{CMB anisotropies generated by cosmic string networks with loops. From left to right, we plot  the TT, TE, EE and BB power spectra, as a function of the multipole moment $\ell$. The top, middle and bottom rows represent the scalar,vector and tensor components, respectively. In each panel, we include the CMB anisotropies generated by a cosmic string network with loops (solid blue line), as well as the contribution of  long strings (dash-dotted black line) and cosmic string loops (dashed red line). The result is obtained by averaging over 500 realizations of cosmic string networks. \label{cmbcor}}
\end{center}
\end{figure}



\begin{figure}
\begin{center}
\includegraphics[width=6.7in]{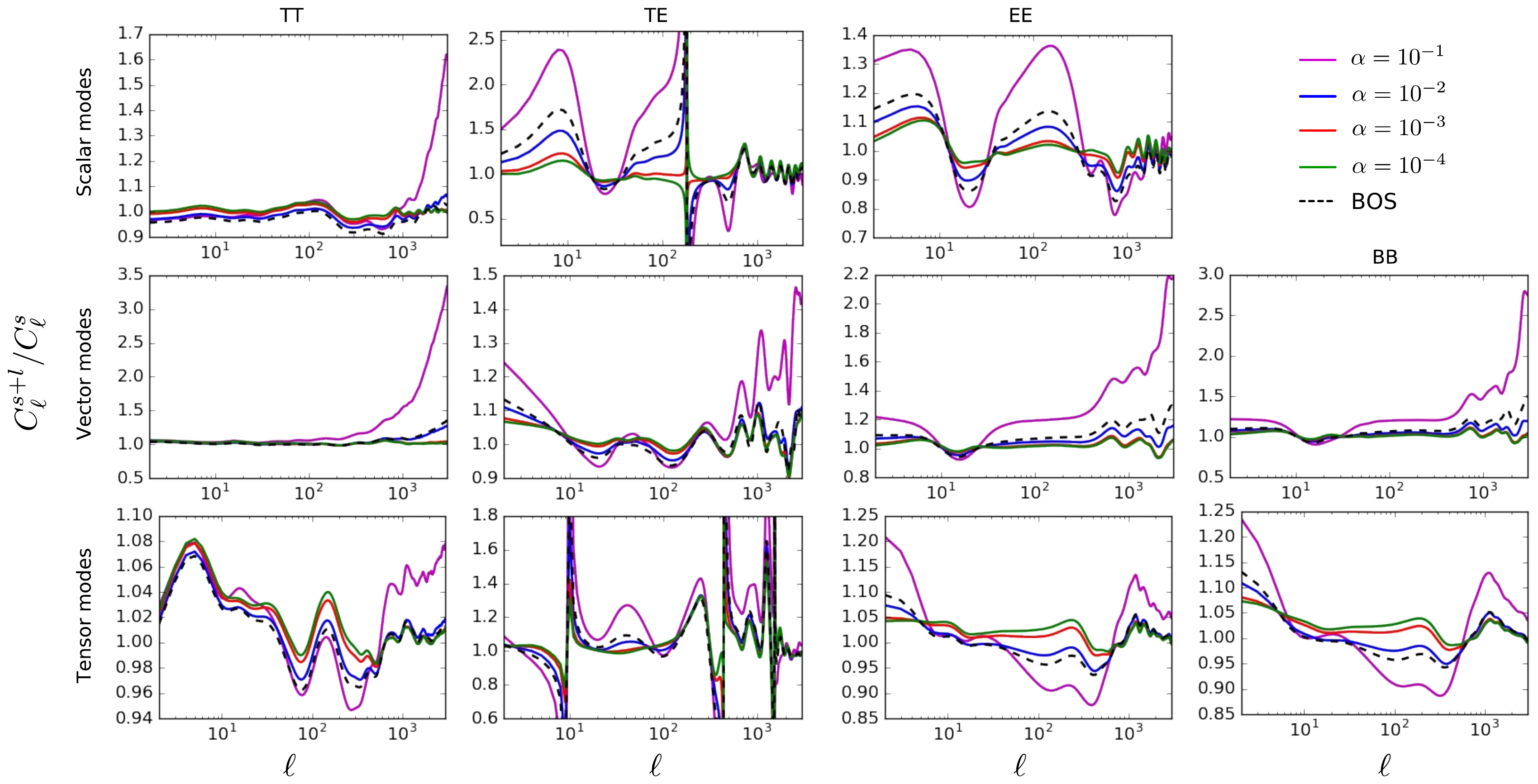}
\caption{Ratio between the CMB anisotropies generated by cosmic string networks with loops, $C^{S+l}_{\ell}$,  and those generated by long strings only, $C^S_{\ell}$, for different loop sizes. From left to right, we plot  the TT, TE, EE and BB power spectra, as a function of the multipole moment $\ell$. The top, middle and bottom rows represent the scalar, vector and tensor components, respectively. Each panel includes the anisotropies generated by networks with $\alpha = 10^{-1}$, $\alpha=10^{-2}$, $\alpha=10^{-3}$, $\alpha=10^{-4}$ and for the loop distribution inferred from the simulations of Blanco-Pillado, Olum and Shlaer (BOS) in Ref.~\cite{Blanco-Pillado:2013qja}.\label{CMBalp}}
\end{center}
\end{figure}

\end{widetext}

\begin{figure}
\begin{center}
\includegraphics[width=3.4in]{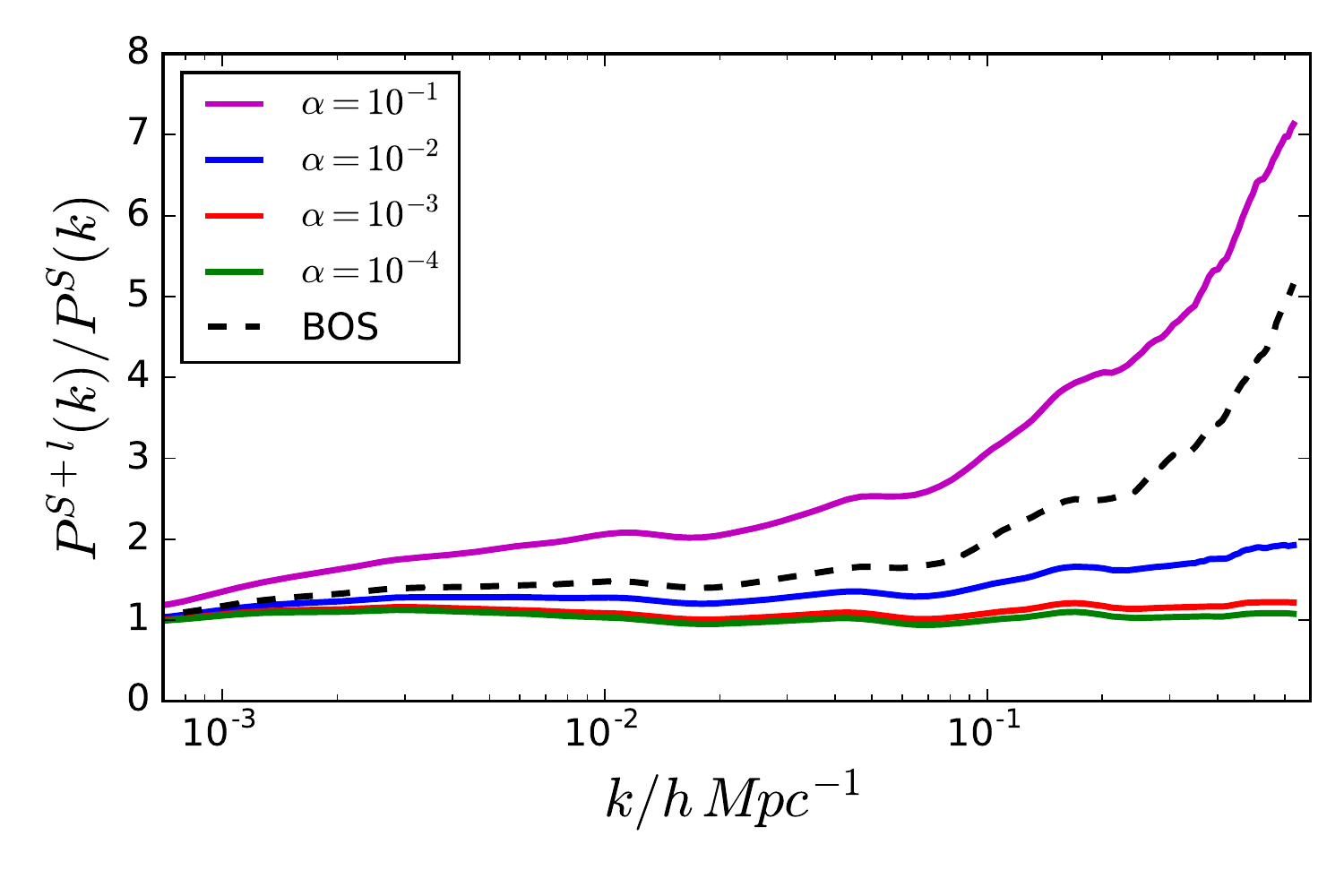}
\caption{Ratio between the linear CDM power spectrum generated by cosmic string networks with loops, $P^{S+l}(k)$, and that of long strings, $P^{S}(k)$, for different loop sizes.  We include the CDM power spectra generated by networks with $\alpha = 10^{-1}$, $\alpha=10^{-2}$, $\alpha=10^{-3}$, $\alpha=10^{-4}$ and for the BOS distribution. \label{pkalp}}
\end{center}
\end{figure}

The CDM linear power spectrum generated by cosmic string networks with loops is plotted in Fig.~\ref{pkcor}, alongside that of cosmic strings and that of (randomly distributed) loops. Therein one may see that, as a result of the correlations, the effect of including loops is to enhance the spectrum of perturbations generated by strings. The shape of the power spectrum is not significantly affected --- except on small scales wherein the decrease is somewhat slighter than $k^{-2}$ due to the inclusion of loops --- but its maximum amplitude increases by a factor of $2-3$. 

The CMB anisotropies generated by a cosmic string network with loops (in both the temperature and polarization channels) are plotted in Fig.~\ref{cmbcor}. In these plots, we may see that loops with $\alpha=10^{-1}$ roughly contribute to about $10\%$ of the anisotropies and lead to a visible increase for large multipole moments $\ell$. This enhancement is particularly significant in the vector  modes --- since, as discussed before, loops generate (due to their shape) vortical motions of matter --- and, in the case of the TT vector anisotropies, there is a significant increase for $l \gtrsim 1000$. Current Planck constraints on Nambu-Goto strings~\cite{Ade:2013xla} --- which limit their fractional contribution to temperature anisotropies to about $1-2\%$ --- were derived using the original USM model and thus do not include the effect of loops. Our results then indicate that the inclusion of loops may result in more stringent constraints on cosmic string tension. Note also that, although the shape of the power spectra generated by cosmic string networks with loops is very similar to that of long strings, the contribution of loops to the vector-mode temperature anisotropies is dominant for large multipole moments $\ell$. As a result, the temperature anisotropies decrease more slowly with increasing $\ell$ if loops are included. This is in agreement with the results of Ref.~\cite{FraisseRingevalSpergelBouchet} --- derived using numerical simulations that include loops ---
in which the TT anisotropies scale as  $\ell^{-0.89}$ for large angular scales (whereas, for the USM model, the TT angular power spectrum behaves asymptotically as $\ell^{-1.5}$\cite{PogosianTyeWassermanWyman2}). Our results indicate that this discrepancy may indeed be related to the contribution of cosmic string loops.

The impact of the inclusion of loops on the anisotropies is highly dependent on the size of loops. Figs.~\ref{CMBalp} and~\ref{pkalp} --- where we plot the TT, TE, EE and BB anisotropies and the linear CDM power spectrum generated by cosmic string networks with loops of different sizes (normalized to that of long strings) --- clearly show this effect. There we also include the spectra generated by Nambu-Goto cosmic string networks with loops calibrated by the latest simulations in Ref.~\cite{Blanco-Pillado:2013qja}. This is done by setting $\alpha\approx 0.34$, $v_l=0.42 $, and by correcting the number of loops produced by a factor of $\mathcal{F}=0.1$ (to account for the fact that only about $10\%$ of the energy lost by network goes into the formation of loops). Our results indicate that, for this scenario, the impact of cosmic string loops on small scales is still significant. Loops then should be considered in the derivation of observational constraints on the tension of Nambu-Goto strings.

Our results show that the contribution of loops remains relevant as the length of the loop decreases. Note however that the correlation between the positions and velocities of long strings and loops play a determinant role here: for randomly distributed loops (as the ones considered in the previous section) the CMB anisotropies die off quickly with decreasing $\alpha$ (roughly $C_\ell\propto\alpha$). However, since loops are distributed along the strings, these loops, even if small, enhance the perturbations that have been generated by the long strings. In fact, as the size of loops decreases, there are two competing effects: the number of loops increases (as $N_l\propto \alpha^{-1}$); but these loops are smaller and survive for a shorter period of time. The net result is that, as these figures show, a decrease of the loops' length leads to a decrease of the impact of loops on both the spectrum of perturbations and CMB anisotropies. The excess in the vector modes in the temperature anisotropies quickly decreases with decreasing $\alpha$ and, as a result, the constraints on scenarios with smaller loops are necessarily less stringent. Interestingly, however, there is an excess in tensor modes even for small $\alpha$, especially in the polarization channels: tensor modes actually increase as $\alpha$ decreases for $10\lesssim \ell\lesssim 10^3$. We have verified numerically that this trend continues if we decrease $\alpha$ further, even beyond the gravitational backreaction scale ($\Gamma G\mu_0$), but the decrease in temperature anisotropies and the increase in polarization anisotropies is significantly slower. In fact, there seems to be a residual excess (similar in magnitude to that observed for $\alpha=10^{-4}$) in both the temperature and polarization channels as a result of the inclusion of loops, even if they are quite small. The B-mode polarization channel may then offer us an alternative window to probe the size of cosmic string loops.

Note that, although here we only plotted the CMB anisotropies generated by cosmic string networks with loops for $G\mu_0=10^{-7}$, we have verified that the contribution of loops to the CMB anisotropies remains relevant as we lower cosmic string tension. As a matter of fact, the amplitude of CMB anisotropies generated by loops scales as $C_\ell\propto (G\mu_0)^2$ (as does that of long strings) and thus the shapes of the (total) angular power spectra are roughly maintained. Note also that, in our computations of the CMB generated by cosmic string networks with loops, we have assumed that $\mathcal{F}=1$ (except for the spectra generated by Nambu-Goto cosmic string networks with loops). If we relax the assumption that all loops are created with the same size and assume thus that $\mathcal{F}\neq 1$, the amplitude of the CMB anisotropies would be suppressed by a factor of $\mathcal{F}^{1/2}$. The specific value of $\mathcal{F}$, however, would depend on the particular distribution assumed for the length of produced loops. If the width of the distribution is not very large, we do not expect this assumption to affect the results significantly, but if there is a large spread this may have an impact on the final results.

\subsection{Strings with reduced intercommutation probability\label{sec:super}}

Until now, we have assumed that the intercommutation probability $P$ is equal to $1$ and thus that whenever cosmic strings collide they exchange partners and reconnect. However, several brane-inflationary scenarios~\cite{DvaliTye, BurgessMajumdarNolteQuevedoRajeshZhang, SarangiTye} predict the production of fundamental strings (or F-strings) and one-dimensional D-branes (or D-strings) that may grow to macroscopic sizes and play the cosmological role of cosmic strings~\cite{JonesStoicaTye, DvaliVilenkin, DvaliKalloshProeyen}. These cosmic superstrings may, unlike ordinary strings, have an intercommutation probability that is significantly reduced due to their quantum nature~\cite{JacksonJonesPolchinski, HananyHashimoto, Jackson_2007} and/or the existence of extra dimensions~\cite{AvgoustidisShellard2}. In this section, we investigate whether the contribution of cosmic string loops to the CMB anisotropies remains significant for string networks with a reduced intercommutation probability. Note however that, although cosmic superstrings serve as the motivation for this study, there are several important properties of these networks that will not be taken into consideration. In particular, since F- and D-strings do not intercommute, but instead bind together to create a new (heavier) type of string, their collision leads to the production of Y-type junctions. Moreover, subsequent collisions are expected to give rise to even heavier $(p,q)$-strings --- composed of $q$ F-strings and $p$ D-strings --- and thus cosmic superstrings are expected to form entangled multi-tensional networks (see e.g Refs.~\cite{TyeWassermanWyman, AvgoustidisShellard3, PourtsidouAvgoustidisCopelandPogosianSteer, RybakAvgoustidisMartins2}). The heavier string types and junctions may have a significant impact on the network's dynamics~\cite{AvgoustidisShellard2,AvgoustidisShellard3} and observational signatures~\cite{PourtsidouAvgoustidisCopelandPogosianSteer,Sousa:2016ggw}. The work in this section should, then, be regarded as exploratory, but our results may indicate us whether it is worth performing a more profund study.

\begin{figure} [h!]
\begin{center}
\includegraphics[width=3.4in]{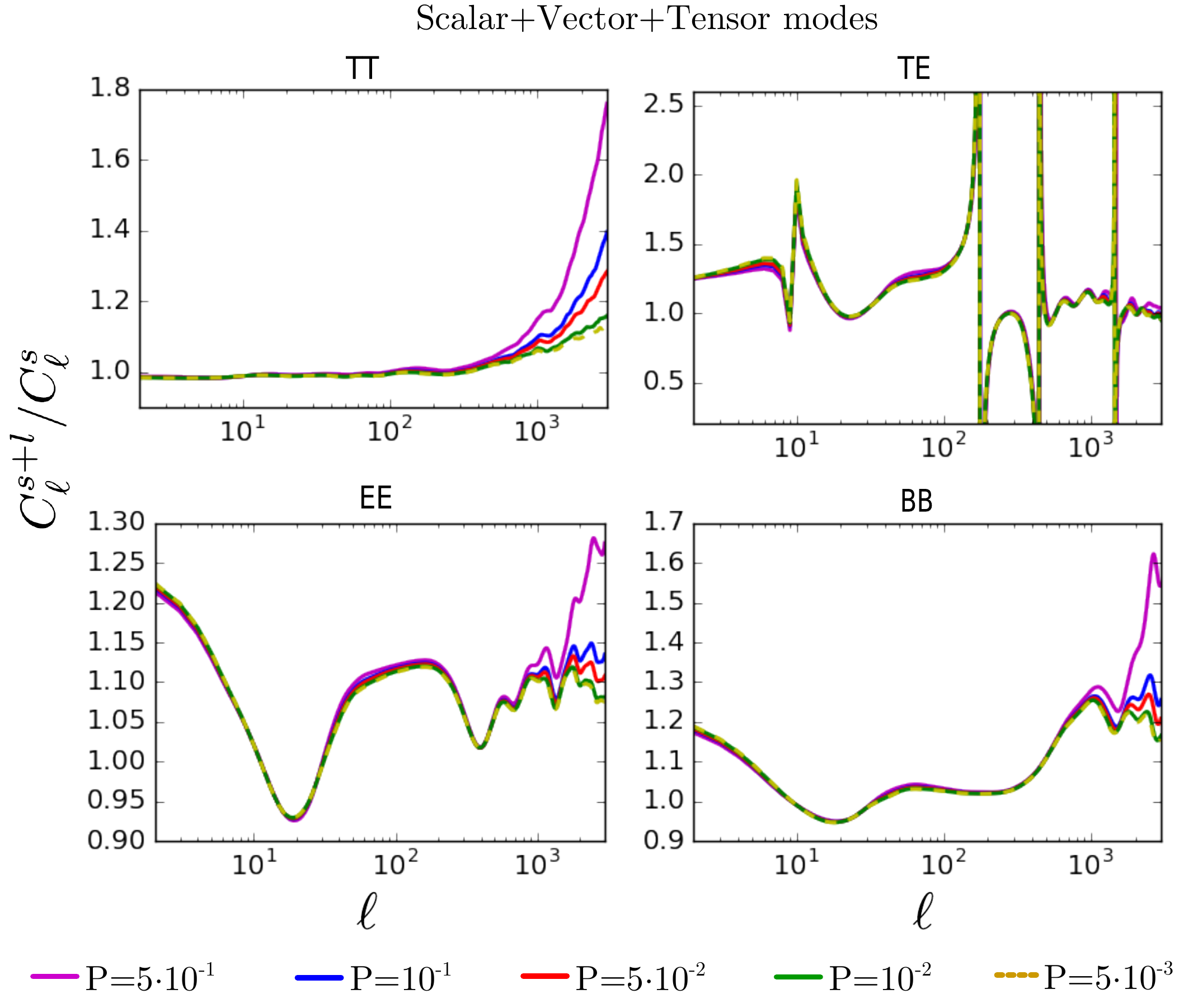}
\caption{Ratio between the CMB anisotropies generated by cosmic string networks with loops, $C^{S+l}_{\ell}$, and those generated by long strings, $C^S_{\ell}$, for different intercommutation probabilities and $\alpha=10^{-1}$. We plot the total TT, TE, EE and BB angular power spectra, as a function of the multipole moment $\ell$. Each panel includes the CMB anisotropies for $P = 5 \cdot 10^{-1}$, $P = 10^{-1}$, $P = 5 \cdot 10^{-2}$, $P = 10^{-2}$, $P = 5 \cdot 10^{-3}$. \label{CMBp}}
\end{center}
\end{figure}

\begin{figure} [h!]
\begin{center}
\includegraphics[width=3.4in]{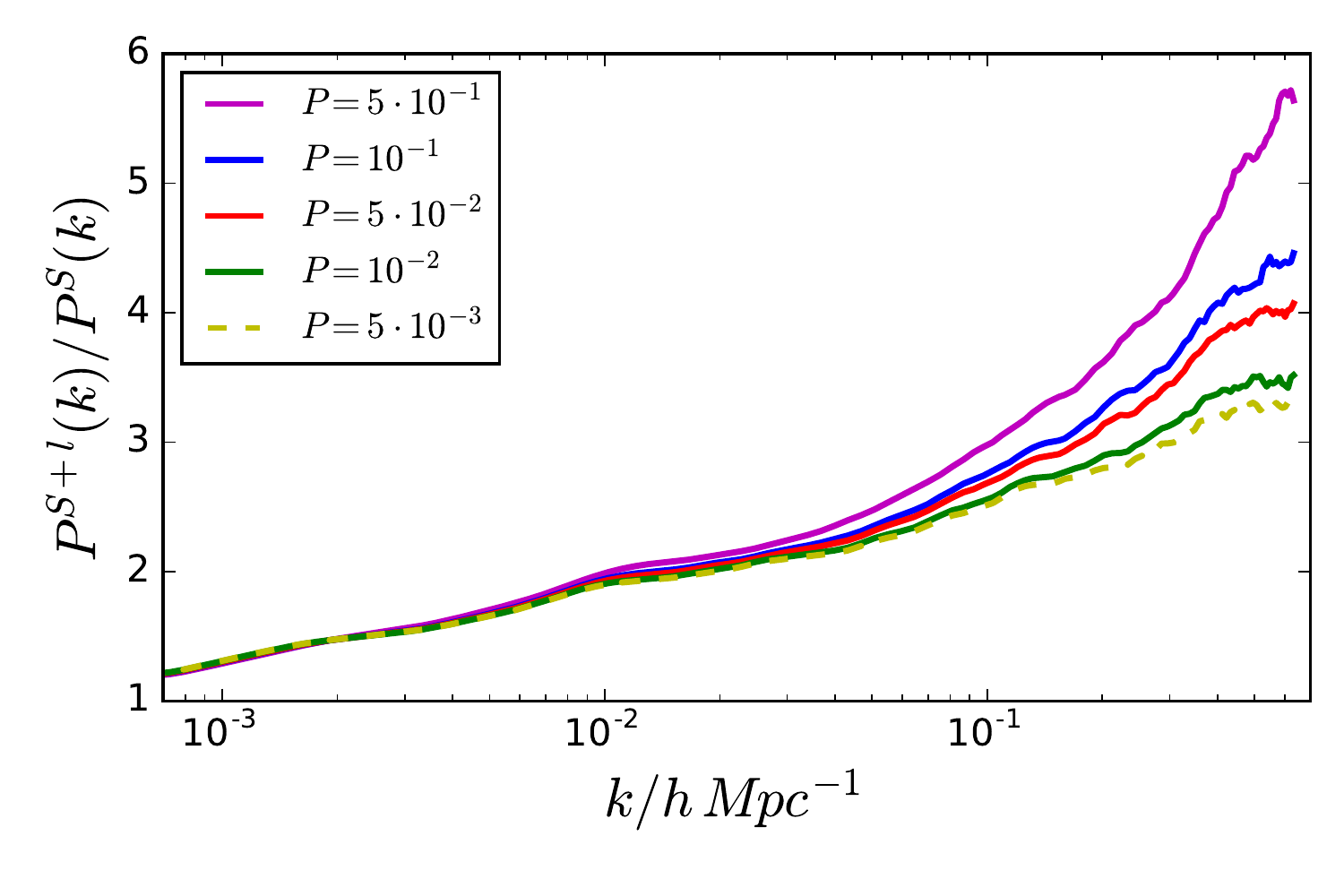}
\caption{Ratio between the linear CDM power spectrum generated by cosmic string networks with loops, $P^{S+l}(k)$, and that generated by long strings, $P^{S}(k)$, for different intercommutation probabilities and $\alpha=10^{-1}$.  We include the spectra generated by networks with $P = 5 \cdot 10^{-1}$, $P = 10^{-1}$, $P = 5 \cdot 10^{-2}$, $P = 10^{-2}$, $P = 5 \cdot 10^{-3}$. \label{PowerP}}
\end{center}
\end{figure}

When two strings with reduced intercommutation probability collide, there is a $1-P$ probability that they merely pass through each other without interaction. This naturally results in a decrease of the loop-chopping efficiency and, consequently, of the network's energy losses. As a matter of fact, radiation and matter era numerical simulations of cosmic string networks in Ref.~\cite{AvgoustidisShellard} with $P<1$ indicate that

\begin{equation}
\label{ProbC}
\tilde{c}(P)= \tilde{c}(1) P^{1/3}\,,
\end{equation}
where $\tilde{c}(1)=0.23$ is the loop-chopping efficiency of networks with $P=1$\footnote{Note however that Minkowski space simulations seem to indicate that $\tilde{c}\propto P^{1/2}$ instead~\cite{Sakellariadou_2005}.}. These networks are, then, weakly interacting and consequently significantly denser, with a characteristic length of $a L_c\sim P^{1/3}$ and $\rho \propto  P ^{-2/3}$~\cite{Avelino:2012qy}. As a result, the energy density lost due to loop formation is actually enhanced as $P$ is reduced:
\be 
\left.\frac{d\rho}{d t}\right|_{loops}\propto P^{-2/3}\,.
\ee
Note however that, although the number of loops created (per unit volume) increases as $P$ decreases, the length of the loops is expected to decrease since the characteristic length of these networks is smaller. The combination of these two effects actually results in a decrease of the contribution of loops to the anisotropies and perturbation spectra. This is illustrated in Figs.~\ref{CMBp} and~\ref{PowerP}, where the CMB and CDM power spectra generated by cosmic string networks with loops for different values of the intercommutation probability are plotted (normalized to those of long strings). Note however that, although the amplitude of the contribution clearly decreases with decreasing $P$, the contribution of loops at small angular scales (large $\ell$) remains significant both in the temperature and polarization channels. Our results indicate that the inclusion of loops in computations of the CMB anisotropies generated by cosmic superstring networks may be necessary to obtain more accurate results. In fact, this contribution may be particularly relevant for the lightest strings (the F-strings), which have a higher intercommutation probability when compared to heavier string types.

\section{Conclusions and outlook \label{sec:conc}}

In this paper, we have studied the impact of loops on the cosmic microwave anisotropies generated by cosmic string networks. To do so, we have extended the USM --- which describes the stress-energy tensor of a network of long strings --- to also account for the contribution of loops and implemented this extended version on the publicly available CMBACT code.

Our results show that cosmic string loops may significantly contribute to the CMB anisotropies on small angular scales (or large multipole moments) on both the temperature and polarization channels. This may lead to more stringent CMB constraints on cosmic string tension for scenarios with larger loop lengths. We further demonstrated that loops with different sizes generate distinct signatures on the polarization angular power spectra. As a result, B-mode polarization may be used to probe different loop-formation scenarios. 

Since loops are expected to decay by emitting gravitational radiation and to give rise to a stochastic gravitational wave background, loop-forming scenarios may also be probed independently using gravitational wave experiments. Current pulsar-timing data set an upper limit for the tension of cosmic strings of about $G\mu_0 \lesssim 10^{-11}$ for large loops~\cite{Lentati:2015qwp,Arzoumanian:2015liz} and Nambu-Goto strings~\cite{Blanco-Pillado:2017rnf}, which is beyond the reach of CMB experiments. Note however that pulsar-timing constraints on scenarios with smaller loops are 3 to 4 orders of magnitude weaker~\cite{Lentati:2015qwp,Arzoumanian:2015liz} and thus the CMB anisotropies may be a competitive independent probe of these scenarios.

Although we have mostly considered standard cosmic string networks in this paper, this framework may be extended to study more exotic scenarios. Here, we have investigated networks with reduced intercommutation probability as a proxy for cosmic superstring networks. Our results indicate that loops may provide a significant contribution in the case of cosmic superstrings too. Note however that a more detailed study, including the contributions of different types of strings and junctions, would be necessary before these results can be safely extrapolated to cosmic superstrings.

This framework may also be helpful to study the CMB signatures of vortons. Axion strings are superconducting and give rise to loops that carry a current~\cite{FukudaManoharMurayamaTelem, AbeHamadaYoshioka}. After creation, these loops decay radiatively, but their current prevents them from evaporating completely. They are, in fact, expected to stabilize after reaching a critical radius and survive throughout cosmological history~\cite{Davis:1988ij, Brandenberger:1996zp}. These stable superconducting loops, known as vortons, have been proposed as possible dark-matter candidates and may lead to specific CMB signatures~\cite{AuclairPeterRingevalSteer}. Note, however, that current is expected to have a significant impact on cosmic string dynamics~\cite{Oliveira:2012nj} and axion string loops are also expected to lose energy at a slower rate. The computation of the signatures of vortons on the CMB would then require an extension of this framework to account for these properties.

\acknowledgments

The authors thank Pedro P. Avelino for many fruitful discussions. L. S. is supported by Funda{\c c}\~ao para a Ci\^encia e a Tecnologia (FCT) through contract No. DL 57/2016/CP1364/CT0001. Funding of this work has also been provided by FCT through national funds (PTDC/FIS-PAR/31938/2017) and by FEDER—Fundo Europeu de Desenvolvimento Regional through COMPETE2020 - Programa Operacional Competitividade e Internacionalização (POCI-01-0145-FEDER-031938), and through the research grants UID/FIS/04434/2019,
UIDB/04434/2020 and UIDP/04434/2020.

\begin{widetext}

\appendix

\section{Stress-energy for loops with $\dot{R_c}\neq 0$} \label{ApA}

It is also possible to obtain a more general expression for loops with $\dot{R_c}\neq 0$. In this case, we have
\begin{equation}
\label{App1}
\textbf{X} = \textbf{X}_0 + R(\gamma_v^{-1} \tau)  \textbf{x} \cos \sigma + R(\gamma_v^{-1} \tau)  \textbf{y} \sin \sigma + v \tau\textbf{z}\,.
\end{equation}
Substituting (\ref{App1}) in the
stress-energy tensor (\ref{StressEnergFour}), we obtain
\begin{equation}
\begin{gathered}
\label{App2}
\Theta^{0 0} = M \gamma_R J_0 (\mathcal{X}) \cos \varphi_0,
\end{gathered}
\end{equation}
where $\gamma_R = (1-\dot{R}^2)^{-1/2}$
and

\begin{equation}
\begin{gathered}
\label{App3}
\Theta^{i j} = \Theta^{0 0}  \left[ v^2 z^i z^j +   \frac{x^i x^j}{2 \gamma_v^2} \left( \dot{R}^2 \mathcal{I}_{+} - \gamma_R^{-2} \mathcal{I}_- \right) + \frac{y^i y^j}{2 \gamma_v^2 } \left( \dot{R}^2 \mathcal{I}_- - \gamma_R^{-2} \mathcal{I}_+ \right) - \mathcal{I} \frac{x^{(i} y^{j)}}{\gamma_v^2} \right] - \\
- 2 M \gamma_R \dot{R} \frac{ v }{\gamma_v}  \left[ x^{(i}z^{j)} \sin B + y^{(i} z^{j)} \cos B \right] J_1(\mathcal{X})   \sin \varphi_0,
\end{gathered}
\end{equation}
The scalar, vector and tensor
contribution from the stress energy 
tensor (\ref{StressEnergFour}) can
be expressed as
\begin{equation}
\begin{gathered}
\label{App4}
\Theta^{S} = \frac{\Theta^{0 0}}{2}  \left[ v^2 (3 z^3 z^3 - 1) + \frac{\gamma_v^{-2}}{2 } \left( (3 x^3 x^3 - 1) (-1+2\dot{R}^2+Y) + (3 y^3 y^3 - 1) (-1+2\dot{R}^2-Y) - 6  \mathcal{I} x^{3} y^{3} \right) \right] - \\
- 2 M \gamma_R \dot{R} \frac{ v }{\gamma_v}  \left[ x^{(i}z^{j)} \sin B + y^{(i} z^{j)} \cos B \right] J_1(\mathcal{X})   \sin \varphi_0, \\
\Theta^{V} = \Theta^{0 0} \left[ v^2 z^1 z^3 + \frac{\gamma_v^{-2}}{2} \left( x^1 x^3 (-1+2 \dot{R}^2 +Y) + y^1 y^3 (-1+2 \dot{R}^2 -Y) - \mathcal{I} (x^1 y^3 + x^3 y^1)  \right) \right] - \\
- M \gamma_R \dot{R} \frac{v}{\gamma_v} \left[ (x^1 z^3 + x^3 z^1) \sin B + (y^1 z^3 + y^3 z^1) \cos B \right] J_1(\mathcal{X}) \sin \varphi_0, \\
\Theta^{T} = \Theta^{0 0} \left[ v^2 z^1 z^2 + \frac{\gamma_v^{-2}}{2} \left( x^1 x^2 (-1+2 \dot{R}^2 +Y) + y^1 y^2 (-1+2 \dot{R}^2 -Y) - \mathcal{I} (x^1 y^2 + x^2 y^1)  \right) \right] - \\
- M \gamma_R \dot{R} \frac{v}{\gamma_v} \left[ (x^1 z^2 + x^2 z^1) \sin B + (y^1 z^2 + y^2 z^1) \cos B \right] J_1(\mathcal{X}) \sin \varphi_0.
\end{gathered}
\end{equation}
\end{widetext}

\bibliography{LoopsCMB}

\begin{thebibliography}{77}%
\makeatletter
\providecommand \@ifxundefined [1]{%
 \@ifx{#1\undefined}
}%
\providecommand \@ifnum [1]{%
 \ifnum #1\expandafter \@firstoftwo
 \else \expandafter \@secondoftwo
 \fi
}%
\providecommand \@ifx [1]{%
 \ifx #1\expandafter \@firstoftwo
 \else \expandafter \@secondoftwo
 \fi
}%
\providecommand \natexlab [1]{#1}%
\providecommand \enquote  [1]{``#1''}%
\providecommand \bibnamefont  [1]{#1}%
\providecommand \bibfnamefont [1]{#1}%
\providecommand \citenamefont [1]{#1}%
\providecommand \href@noop [0]{\@secondoftwo}%
\providecommand \href [0]{\begingroup \@sanitize@url \@href}%
\providecommand \@href[1]{\@@startlink{#1}\@@href}%
\providecommand \@@href[1]{\endgroup#1\@@endlink}%
\providecommand \@sanitize@url [0]{\catcode `\\12\catcode `\$12\catcode
  `\&12\catcode `\#12\catcode `\^12\catcode `\_12\catcode `\%12\relax}%
\providecommand \@@startlink[1]{}%
\providecommand \@@endlink[0]{}%
\providecommand \url  [0]{\begingroup\@sanitize@url \@url }%
\providecommand \@url [1]{\endgroup\@href {#1}{\urlprefix }}%
\providecommand \urlprefix  [0]{URL }%
\providecommand \Eprint [0]{\href }%
\providecommand \doibase [0]{https://doi.org/}%
\providecommand \selectlanguage [0]{\@gobble}%
\providecommand \bibinfo  [0]{\@secondoftwo}%
\providecommand \bibfield  [0]{\@secondoftwo}%
\providecommand \translation [1]{[#1]}%
\providecommand \BibitemOpen [0]{}%
\providecommand \bibitemStop [0]{}%
\providecommand \bibitemNoStop [0]{.\EOS\space}%
\providecommand \EOS [0]{\spacefactor3000\relax}%
\providecommand \BibitemShut  [1]{\csname bibitem#1\endcsname}%
\let\auto@bib@innerbib\@empty
\bibitem [{\citenamefont {Hindmarsh}\ and\ \citenamefont
  {Kibble}(1995)}]{HindmarshKibble}%
  \BibitemOpen
  \bibfield  {author} {\bibinfo {author} {\bibfnamefont {M.~B.}\ \bibnamefont
  {Hindmarsh}}\ and\ \bibinfo {author} {\bibfnamefont {T.~W.~B.}\ \bibnamefont
  {Kibble}},\ }\bibfield  {title} {\bibinfo {title} {Cosmic strings},\ }\href
  {https://doi.org/10.1088/0034-4885/58/5/001} {\bibfield  {journal} {\bibinfo
  {journal} {Rept.Prog.Phys.}\ }\textbf {\bibinfo {volume} {58}},\ \bibinfo
  {pages} {477} (\bibinfo {year} {1995})},\ \Eprint
  {https://arxiv.org/abs/hep-ph/9411342v1} {arXiv:hep-ph/9411342v1
  [astro-ph.CO]} \BibitemShut {NoStop}%
\bibitem [{\citenamefont {Vilenkin}\ and\ \citenamefont
  {Shellard}(2000)}]{ShellardVilenkin}%
  \BibitemOpen
  \bibfield  {author} {\bibinfo {author} {\bibfnamefont {A.}~\bibnamefont
  {Vilenkin}}\ and\ \bibinfo {author} {\bibfnamefont {E.~P.~S.}\ \bibnamefont
  {Shellard}},\ }\href@noop {} {\emph {\bibinfo {title} {Cosmic Strings and
  Other Topological Defects}}}\ (\bibinfo  {publisher} {Cambridge University
  Press, Cambridge},\ \bibinfo {year} {2000})\BibitemShut {NoStop}%
\bibitem [{\citenamefont {Copeland}\ and\ \citenamefont
  {Kibble}(2010)}]{CopelandKibble}%
  \BibitemOpen
  \bibfield  {author} {\bibinfo {author} {\bibfnamefont {E.~J.}\ \bibnamefont
  {Copeland}}\ and\ \bibinfo {author} {\bibfnamefont {T.~W.~B.}\ \bibnamefont
  {Kibble}},\ }\bibfield  {title} {\bibinfo {title} {Cosmic strings and
  superstrings},\ }\href {https://doi.org/10.1098/rspa.2009.0591} {\bibfield
  {journal} {\bibinfo  {journal} {Proceedings of the Royal Society A:
  Mathematical, Physical and Engineering Sciences}\ }\textbf {\bibinfo {volume}
  {466}},\ \bibinfo {pages} {623} (\bibinfo {year} {2010})},\ \Eprint
  {https://arxiv.org/abs/0911.1345v3} {arXiv:0911.1345v3 [astro-ph.CO]}
  \BibitemShut {NoStop}%
\bibitem [{\citenamefont {Fixsen}\ \emph {et~al.}(1996)\citenamefont {Fixsen},
  \citenamefont {Cheng}, \citenamefont {Gales}, \citenamefont {Mather},
  \citenamefont {Shafer},\ and\ \citenamefont {Wright}}]{Fixsen:1996nj}%
  \BibitemOpen
  \bibfield  {author} {\bibinfo {author} {\bibfnamefont {D.~J.}\ \bibnamefont
  {Fixsen}}, \bibinfo {author} {\bibfnamefont {E.~S.}\ \bibnamefont {Cheng}},
  \bibinfo {author} {\bibfnamefont {J.~M.}\ \bibnamefont {Gales}}, \bibinfo
  {author} {\bibfnamefont {J.~C.}\ \bibnamefont {Mather}}, \bibinfo {author}
  {\bibfnamefont {R.~A.}\ \bibnamefont {Shafer}},\ and\ \bibinfo {author}
  {\bibfnamefont {E.~L.}\ \bibnamefont {Wright}},\ }\bibfield  {title}
  {\bibinfo {title} {{The Cosmic Microwave Background spectrum from the full
  COBE FIRAS data set}},\ }\href {https://doi.org/10.1086/178173} {\bibfield
  {journal} {\bibinfo  {journal} {Astrophys. J.}\ }\textbf {\bibinfo {volume}
  {473}},\ \bibinfo {pages} {576} (\bibinfo {year} {1996})},\ \Eprint
  {https://arxiv.org/abs/astro-ph/9605054} {arXiv:astro-ph/9605054}
  \BibitemShut {NoStop}%
\bibitem [{\citenamefont {{{Planck Collaboration XXV}}}(2014)}]{Planck2013}%
  \BibitemOpen
  \bibfield  {author} {\bibinfo {author} {\bibnamefont {{{Planck Collaboration
  XXV}}}} (\bibinfo {collaboration} {Planck}),\ }\bibfield  {title} {\bibinfo
  {title} {{Planck 2013 results. {XXV}. Searches for cosmic strings and other
  topological defects}},\ }\href {https://doi.org/10.1051/0004-6361/201321621}
  {\bibfield  {journal} {\bibinfo  {journal} {Astron. Astrophys.}\ }\textbf
  {\bibinfo {volume} {571}},\ \bibinfo {pages} {A25} (\bibinfo {year}
  {2014})},\ \Eprint {https://arxiv.org/abs/arXiv:1303.5085v1}
  {arXiv:arXiv:1303.5085v1 [astro-ph.CO]} \BibitemShut {NoStop}%
\bibitem [{\citenamefont {Lizarraga}\ \emph {et~al.}(2016)\citenamefont
  {Lizarraga}, \citenamefont {Urrestilla}, \citenamefont {Daverio},
  \citenamefont {Hindmarsh},\ and\ \citenamefont
  {Kunz}}]{LizarragaUrrestillaDaverioHindmarshKunz}%
  \BibitemOpen
  \bibfield  {author} {\bibinfo {author} {\bibfnamefont {J.}~\bibnamefont
  {Lizarraga}}, \bibinfo {author} {\bibfnamefont {J.}~\bibnamefont
  {Urrestilla}}, \bibinfo {author} {\bibfnamefont {D.}~\bibnamefont {Daverio}},
  \bibinfo {author} {\bibfnamefont {M.}~\bibnamefont {Hindmarsh}},\ and\
  \bibinfo {author} {\bibfnamefont {M.}~\bibnamefont {Kunz}},\ }\bibfield
  {title} {\bibinfo {title} {New {CMB} constraints for {A}belian {H}iggs cosmic
  strings},\ }\href {https://doi.org/10.1088/1475-7516/2016/10/042} {\bibfield
  {journal} {\bibinfo  {journal} {JCAP}\ }\textbf {\bibinfo {volume}
  {1610}}\bibfield  {number} {\bibinfo  {number} { (10)},\ \bibinfo {pages}
  {042}},\ }\Eprint {https://arxiv.org/abs/1609.03386v3} {arXiv:1609.03386v3
  [astro-ph.CO]} \BibitemShut {NoStop}%
\bibitem [{\citenamefont {Lazanu}\ \emph {et~al.}(2015)\citenamefont {Lazanu},
  \citenamefont {Shellard},\ and\ \citenamefont {Landriau}}]{LazanauShellard1}%
  \BibitemOpen
  \bibfield  {author} {\bibinfo {author} {\bibfnamefont {A.}~\bibnamefont
  {Lazanu}}, \bibinfo {author} {\bibfnamefont {E.~P.~S.}\ \bibnamefont
  {Shellard}},\ and\ \bibinfo {author} {\bibfnamefont {M.}~\bibnamefont
  {Landriau}},\ }\bibfield  {title} {\bibinfo {title} {Cmb power spectrum of
  nambu-goto cosmic strings},\ }\href
  {https://doi.org/10.1103/PhysRevD.91.083519} {\bibfield  {journal} {\bibinfo
  {journal} {Phys. Rev. D}\ }\textbf {\bibinfo {volume} {91}},\ \bibinfo
  {pages} {083519} (\bibinfo {year} {2015})},\ \Eprint
  {https://arxiv.org/abs/1410.4860} {arXiv:1410.4860 [astro-ph.CO]}
  \BibitemShut {NoStop}%
\bibitem [{\citenamefont {Lazanu}\ and\ \citenamefont
  {Shellard}(2015)}]{LazanauShellard}%
  \BibitemOpen
  \bibfield  {author} {\bibinfo {author} {\bibfnamefont {A.}~\bibnamefont
  {Lazanu}}\ and\ \bibinfo {author} {\bibfnamefont {E.~P.~S.}\ \bibnamefont
  {Shellard}},\ }\bibfield  {title} {\bibinfo {title} {Constraints on the
  nambu-goto cosmic string contribution to the cmb power spectrum in light of
  new temperature and polarisation data},\ }\href
  {https://doi.org/10.1088/1475-7516/2015/02/024} {\bibfield  {journal}
  {\bibinfo  {journal} {JCAP}\ }\textbf {\bibinfo {volume} {2015}}\bibfield
  {number} {\bibinfo  {number} { (02)},\ \bibinfo {pages} {024}},\ }\Eprint
  {https://arxiv.org/abs/1410.5046v3} {arXiv:1410.5046v3 [astro-ph.CO]}
  \BibitemShut {NoStop}%
\bibitem [{\citenamefont {Charnock}\ \emph {et~al.}(2016)\citenamefont
  {Charnock}, \citenamefont {Avgoustidis}, \citenamefont {Copeland},\ and\
  \citenamefont {Moss}}]{CharnockAvgoustidisCopelandMoss}%
  \BibitemOpen
  \bibfield  {author} {\bibinfo {author} {\bibfnamefont {T.}~\bibnamefont
  {Charnock}}, \bibinfo {author} {\bibfnamefont {A.}~\bibnamefont
  {Avgoustidis}}, \bibinfo {author} {\bibfnamefont {E.~J.}\ \bibnamefont
  {Copeland}},\ and\ \bibinfo {author} {\bibfnamefont {A.}~\bibnamefont
  {Moss}},\ }\bibfield  {title} {\bibinfo {title} {Cmb constraints on cosmic
  strings and superstrings},\ }\href
  {https://doi.org/10.1103/PhysRevD.93.123503} {\bibfield  {journal} {\bibinfo
  {journal} {Phys. Rev. D}\ }\textbf {\bibinfo {volume} {93}},\ \bibinfo
  {pages} {123503} (\bibinfo {year} {2016})},\ \Eprint
  {https://arxiv.org/abs/1603.01275} {arXiv:1603.01275 [astro-ph.CO]}
  \BibitemShut {NoStop}%
\bibitem [{\citenamefont {Fraisse}\ \emph {et~al.}(2008)\citenamefont
  {Fraisse}, \citenamefont {Ringeval}, \citenamefont {Spergel},\ and\
  \citenamefont {Bouchet}}]{FraisseRingevalSpergelBouchet}%
  \BibitemOpen
  \bibfield  {author} {\bibinfo {author} {\bibfnamefont {A.~A.}\ \bibnamefont
  {Fraisse}}, \bibinfo {author} {\bibfnamefont {C.}~\bibnamefont {Ringeval}},
  \bibinfo {author} {\bibfnamefont {D.~N.}\ \bibnamefont {Spergel}},\ and\
  \bibinfo {author} {\bibfnamefont {F.~m. c.~R.}\ \bibnamefont {Bouchet}},\
  }\bibfield  {title} {\bibinfo {title} {Small-angle cmb temperature
  anisotropies induced by cosmic strings},\ }\href
  {https://doi.org/10.1103/PhysRevD.78.043535} {\bibfield  {journal} {\bibinfo
  {journal} {Phys. Rev. D}\ }\textbf {\bibinfo {volume} {78}},\ \bibinfo
  {pages} {043535} (\bibinfo {year} {2008})},\ \Eprint
  {https://arxiv.org/abs/0708.1162} {arXiv:0708.1162 [astro-ph]} \BibitemShut
  {NoStop}%
\bibitem [{\citenamefont {Brandenberger}\ and\ \citenamefont
  {Turok}(1986)}]{BrandenbergerTurok}%
  \BibitemOpen
  \bibfield  {author} {\bibinfo {author} {\bibfnamefont {R.~H.}\ \bibnamefont
  {Brandenberger}}\ and\ \bibinfo {author} {\bibfnamefont {N.}~\bibnamefont
  {Turok}},\ }\bibfield  {title} {\bibinfo {title} {Fluctuations from cosmic
  strings and the microwave background},\ }\href
  {https://doi.org/10.1103/PhysRevD.33.2182} {\bibfield  {journal} {\bibinfo
  {journal} {Phys. Rev. D}\ }\textbf {\bibinfo {volume} {33}},\ \bibinfo
  {pages} {2182} (\bibinfo {year} {1986})}\BibitemShut {NoStop}%
\bibitem [{\citenamefont {Traschen}\ \emph {et~al.}(1986)\citenamefont
  {Traschen}, \citenamefont {Turok},\ and\ \citenamefont
  {Brandenberger}}]{TraschenTurokBrandenberger}%
  \BibitemOpen
  \bibfield  {author} {\bibinfo {author} {\bibfnamefont {J.}~\bibnamefont
  {Traschen}}, \bibinfo {author} {\bibfnamefont {N.}~\bibnamefont {Turok}},\
  and\ \bibinfo {author} {\bibfnamefont {R.}~\bibnamefont {Brandenberger}},\
  }\bibfield  {title} {\bibinfo {title} {Microwave anisotropies from cosmic
  strings},\ }\href {https://doi.org/10.1103/PhysRevD.34.919} {\bibfield
  {journal} {\bibinfo  {journal} {Phys. Rev. D}\ }\textbf {\bibinfo {volume}
  {34}},\ \bibinfo {pages} {919} (\bibinfo {year} {1986})}\BibitemShut
  {NoStop}%
\bibitem [{\citenamefont {Pogosian}\ and\ \citenamefont
  {Vachaspati}(1999)}]{PogosianVachaspati}%
  \BibitemOpen
  \bibfield  {author} {\bibinfo {author} {\bibfnamefont {L.}~\bibnamefont
  {Pogosian}}\ and\ \bibinfo {author} {\bibfnamefont {T.}~\bibnamefont
  {Vachaspati}},\ }\bibfield  {title} {\bibinfo {title} {Cosmic microwave
  background anisotropy from wiggly strings},\ }\href
  {https://doi.org/10.1103/PhysRevD.60.083504} {\bibfield  {journal} {\bibinfo
  {journal} {Phys.Rev.}\ }\textbf {\bibinfo {volume} {D60}},\ \bibinfo {pages}
  {083504} (\bibinfo {year} {1999})},\ \Eprint
  {https://arxiv.org/abs/astro-ph/9903361v4} {arXiv:astro-ph/9903361v4
  [astro-ph.CO]} \BibitemShut {NoStop}%
\bibitem [{\citenamefont {Albrecht}\ \emph {et~al.}(1998)\citenamefont
  {Albrecht}, \citenamefont {Battye},\ and\ \citenamefont
  {Robinson}}]{AlbrechtBattyeRobinson}%
  \BibitemOpen
  \bibfield  {author} {\bibinfo {author} {\bibfnamefont {A.}~\bibnamefont
  {Albrecht}}, \bibinfo {author} {\bibfnamefont {R.~A.}\ \bibnamefont
  {Battye}},\ and\ \bibinfo {author} {\bibfnamefont {J.}~\bibnamefont
  {Robinson}},\ }\bibfield  {title} {\bibinfo {title} {Detailed study of defect
  models for cosmic structure formation},\ }\href
  {https://doi.org/10.1103/PhysRevD.59.023508} {\bibfield  {journal} {\bibinfo
  {journal} {Phys. Rev. D}\ }\textbf {\bibinfo {volume} {59}},\ \bibinfo
  {pages} {023508} (\bibinfo {year} {1998})},\ \Eprint
  {https://arxiv.org/abs/astro-ph/9711121v2} {arXiv:astro-ph/9711121v2
  [astro-ph]} \BibitemShut {NoStop}%
\bibitem [{\citenamefont {Rybak}\ \emph {et~al.}(2017)\citenamefont {Rybak},
  \citenamefont {Avgoustidis},\ and\ \citenamefont
  {Martins}}]{RybakAvgoustidisMartins}%
  \BibitemOpen
  \bibfield  {author} {\bibinfo {author} {\bibfnamefont {I.~{\relax Yu}.}\
  \bibnamefont {Rybak}}, \bibinfo {author} {\bibfnamefont {A.}~\bibnamefont
  {Avgoustidis}},\ and\ \bibinfo {author} {\bibfnamefont {C.~J. A.~P.}\
  \bibnamefont {Martins}},\ }\bibfield  {title} {\bibinfo {title}
  {{Semianalytic calculation of cosmic microwave background anisotropies from
  wiggly and superconducting cosmic strings}},\ }\href
  {https://doi.org/10.1103/PhysRevD.96.103535, 10.1103/PhysRevD.100.049901}
  {\bibfield  {journal} {\bibinfo  {journal} {Phys. Rev.}\ }\textbf {\bibinfo
  {volume} {D96}},\ \bibinfo {pages} {103535} (\bibinfo {year} {2017})},\
  \bibinfo {note} {[Erratum: Phys. Rev.D100,no.4,049901(2019)]},\ \Eprint
  {https://arxiv.org/abs/1709.01839} {arXiv:1709.01839 [astro-ph.CO]}
  \BibitemShut {NoStop}%
\bibitem [{\citenamefont {Sousa}\ and\ \citenamefont
  {Avelino}(2015)}]{SousaAvelino}%
  \BibitemOpen
  \bibfield  {author} {\bibinfo {author} {\bibfnamefont {L.}~\bibnamefont
  {Sousa}}\ and\ \bibinfo {author} {\bibfnamefont {P.~P.}\ \bibnamefont
  {Avelino}},\ }\bibfield  {title} {\bibinfo {title} {Cosmic microwave
  background anisotropies generated by domain wall networks},\ }\href
  {https://doi.org/10.1103/PhysRevD.92.083520} {\bibfield  {journal} {\bibinfo
  {journal} {Phys. Rev. D}\ }\textbf {\bibinfo {volume} {92}},\ \bibinfo
  {pages} {083520} (\bibinfo {year} {2015})},\ \Eprint
  {https://arxiv.org/abs/1507.01064v1} {arXiv:1507.01064v1 [astro-ph.CO]}
  \BibitemShut {NoStop}%
\bibitem [{\citenamefont {Avgoustidis}\ \emph {et~al.}(2012)\citenamefont
  {Avgoustidis}, \citenamefont {Copeland}, \citenamefont {Moss},\ and\
  \citenamefont {Skliros}}]{AvgoustidisCopeland}%
  \BibitemOpen
  \bibfield  {author} {\bibinfo {author} {\bibfnamefont {A.}~\bibnamefont
  {Avgoustidis}}, \bibinfo {author} {\bibfnamefont {E.~J.}\ \bibnamefont
  {Copeland}}, \bibinfo {author} {\bibfnamefont {A.}~\bibnamefont {Moss}},\
  and\ \bibinfo {author} {\bibfnamefont {D.}~\bibnamefont {Skliros}},\
  }\bibfield  {title} {\bibinfo {title} {Fast analytic computation of cosmic
  string power spectra},\ }\href {https://doi.org/10.1103/PhysRevD.86.123513}
  {\bibfield  {journal} {\bibinfo  {journal} {Phys. Rev. D}\ }\textbf {\bibinfo
  {volume} {86}},\ \bibinfo {pages} {123513} (\bibinfo {year} {2012})},\
  \Eprint {https://arxiv.org/abs/1209.2461v2} {arXiv:1209.2461v2 [astro-ph.CO]}
  \BibitemShut {NoStop}%
\bibitem [{\citenamefont {Wu}\ \emph {et~al.}(2002)\citenamefont {Wu},
  \citenamefont {Avelino}, \citenamefont {Shellard},\ and\ \citenamefont
  {Allen}}]{Wu:1998mr}%
  \BibitemOpen
  \bibfield  {author} {\bibinfo {author} {\bibfnamefont {J.~H.~P.}\
  \bibnamefont {Wu}}, \bibinfo {author} {\bibfnamefont {P.~P.}\ \bibnamefont
  {Avelino}}, \bibinfo {author} {\bibfnamefont {E.~P.~S.}\ \bibnamefont
  {Shellard}},\ and\ \bibinfo {author} {\bibfnamefont {B.}~\bibnamefont
  {Allen}},\ }\bibfield  {title} {\bibinfo {title} {{Cosmic strings, loops, and
  linear growth of matter perturbations}},\ }\href
  {https://doi.org/10.1142/S0218271802001299} {\bibfield  {journal} {\bibinfo
  {journal} {Int. J. Mod. Phys. D}\ }\textbf {\bibinfo {volume} {11}},\
  \bibinfo {pages} {61} (\bibinfo {year} {2002})},\ \Eprint
  {https://arxiv.org/abs/astro-ph/9812156} {arXiv:astro-ph/9812156}
  \BibitemShut {NoStop}%
\bibitem [{\citenamefont {Moore}\ \emph {et~al.}(2001)\citenamefont {Moore},
  \citenamefont {Shellard},\ and\ \citenamefont
  {Martins}}]{MooreShellardMartins}%
  \BibitemOpen
  \bibfield  {author} {\bibinfo {author} {\bibfnamefont {J.~N.}\ \bibnamefont
  {Moore}}, \bibinfo {author} {\bibfnamefont {E.~P.~S.}\ \bibnamefont
  {Shellard}},\ and\ \bibinfo {author} {\bibfnamefont {C.~J. A.~P.}\
  \bibnamefont {Martins}},\ }\bibfield  {title} {\bibinfo {title} {Evolution of
  abelian-higgs string networks},\ }\href
  {https://doi.org/10.1103/PhysRevD.65.023503} {\bibfield  {journal} {\bibinfo
  {journal} {Phys. Rev. D}\ }\textbf {\bibinfo {volume} {65}},\ \bibinfo
  {pages} {023503} (\bibinfo {year} {2001})},\ \Eprint
  {https://arxiv.org/abs/hep-ph/0107171} {arXiv:hep-ph/0107171 [hep-ph]}
  \BibitemShut {NoStop}%
\bibitem [{\citenamefont {Martins}\ and\ \citenamefont
  {Shellard}(2006)}]{MartinsShellard2006}%
  \BibitemOpen
  \bibfield  {author} {\bibinfo {author} {\bibfnamefont {C.~J. A.~P.}\
  \bibnamefont {Martins}}\ and\ \bibinfo {author} {\bibfnamefont {E.~P.~S.}\
  \bibnamefont {Shellard}},\ }\bibfield  {title} {\bibinfo {title} {Fractal
  properties and small-scale structure of cosmic string networks},\ }\href
  {https://doi.org/10.1103/PhysRevD.73.043515} {\bibfield  {journal} {\bibinfo
  {journal} {Phys. Rev. D}\ }\textbf {\bibinfo {volume} {73}},\ \bibinfo
  {pages} {043515} (\bibinfo {year} {2006})},\ \Eprint
  {https://arxiv.org/abs/astro-ph/0511792} {arXiv:astro-ph/0511792 [astro-ph]}
  \BibitemShut {NoStop}%
\bibitem [{\citenamefont {Blanco-Pillado}\ \emph {et~al.}(2011)\citenamefont
  {Blanco-Pillado}, \citenamefont {Olum},\ and\ \citenamefont
  {Shlaer}}]{BlancoPilladoOlumShlaer}%
  \BibitemOpen
  \bibfield  {author} {\bibinfo {author} {\bibfnamefont {J.~J.}\ \bibnamefont
  {Blanco-Pillado}}, \bibinfo {author} {\bibfnamefont {K.~D.}\ \bibnamefont
  {Olum}},\ and\ \bibinfo {author} {\bibfnamefont {B.}~\bibnamefont {Shlaer}},\
  }\bibfield  {title} {\bibinfo {title} {Large parallel cosmic string
  simulations: New results on loop production},\ }\href
  {https://doi.org/10.1103/PhysRevD.83.083514} {\bibfield  {journal} {\bibinfo
  {journal} {Phys. Rev. D}\ }\textbf {\bibinfo {volume} {83}},\ \bibinfo
  {pages} {083514} (\bibinfo {year} {2011})},\ \Eprint
  {https://arxiv.org/abs/1101.5173} {arXiv:1101.5173 [astro-ph.CO]}
  \BibitemShut {NoStop}%
\bibitem [{\citenamefont {Hindmarsh}\ \emph {et~al.}(2017)\citenamefont
  {Hindmarsh}, \citenamefont {Lizarraga}, \citenamefont {Urrestilla},
  \citenamefont {Daverio},\ and\ \citenamefont
  {Kunz}}]{HindmarshLizarragaUrrestillaDaverioKunz}%
  \BibitemOpen
  \bibfield  {author} {\bibinfo {author} {\bibfnamefont {M.}~\bibnamefont
  {Hindmarsh}}, \bibinfo {author} {\bibfnamefont {J.}~\bibnamefont
  {Lizarraga}}, \bibinfo {author} {\bibfnamefont {J.}~\bibnamefont
  {Urrestilla}}, \bibinfo {author} {\bibfnamefont {D.}~\bibnamefont
  {Daverio}},\ and\ \bibinfo {author} {\bibfnamefont {M.}~\bibnamefont
  {Kunz}},\ }\bibfield  {title} {\bibinfo {title} {Scaling from gauge and
  scalar radiation in abelian-higgs string networks},\ }\href
  {https://doi.org/10.1103/PhysRevD.96.023525} {\bibfield  {journal} {\bibinfo
  {journal} {Phys. Rev. D}\ }\textbf {\bibinfo {volume} {96}},\ \bibinfo
  {pages} {023525} (\bibinfo {year} {2017})},\ \Eprint
  {https://arxiv.org/abs/1703.06696} {arXiv:1703.06696 [astro-ph.CO]}
  \BibitemShut {NoStop}%
\bibitem [{\citenamefont {Correia}\ and\ \citenamefont
  {Martins}(2021)}]{CorreiaMartins2}%
  \BibitemOpen
  \bibfield  {author} {\bibinfo {author} {\bibfnamefont {J.}~\bibnamefont
  {Correia}}\ and\ \bibinfo {author} {\bibfnamefont {C.}~\bibnamefont
  {Martins}},\ }\bibfield  {title} {\bibinfo {title} {Abelian-higgs cosmic
  string evolution with multiple gpus},\ }\href
  {https://doi.org/https://doi.org/10.1016/j.ascom.2020.100438} {\bibfield
  {journal} {\bibinfo  {journal} {Astronomy and Computing}\ }\textbf {\bibinfo
  {volume} {34}},\ \bibinfo {pages} {100438} (\bibinfo {year} {2021})},\
  \Eprint {https://arxiv.org/abs/2005.14454} {arXiv:2005.14454
  [physics.comp-ph]} \BibitemShut {NoStop}%
\bibitem [{\citenamefont {Martins}\ and\ \citenamefont
  {Shellard}(1996)}]{MartinsShellard}%
  \BibitemOpen
  \bibfield  {author} {\bibinfo {author} {\bibfnamefont {C.~J. A.~P.}\
  \bibnamefont {Martins}}\ and\ \bibinfo {author} {\bibfnamefont {E.~P.~S.}\
  \bibnamefont {Shellard}},\ }\bibfield  {title} {\bibinfo {title}
  {Quantitative string evolution},\ }\href
  {https://doi.org/10.1103/PhysRevD.54.2535} {\bibfield  {journal} {\bibinfo
  {journal} {Phys. Rev. D}\ }\textbf {\bibinfo {volume} {54}},\ \bibinfo
  {pages} {2535} (\bibinfo {year} {1996})},\ \Eprint
  {https://arxiv.org/abs/hep-ph/9602271v2} {arXiv:hep-ph/9602271v2 [hep-ph]}
  \BibitemShut {NoStop}%
\bibitem [{\citenamefont {Martins}\ and\ \citenamefont
  {Shellard}(2002)}]{MartinsShellard2}%
  \BibitemOpen
  \bibfield  {author} {\bibinfo {author} {\bibfnamefont {C.~J. A.~P.}\
  \bibnamefont {Martins}}\ and\ \bibinfo {author} {\bibfnamefont {E.~P.~S.}\
  \bibnamefont {Shellard}},\ }\bibfield  {title} {\bibinfo {title} {Extending
  the velocity-dependent one-scale string evolution model},\ }\href
  {https://doi.org/10.1103/PhysRevD.65.043514} {\bibfield  {journal} {\bibinfo
  {journal} {Phys. Rev. D}\ }\textbf {\bibinfo {volume} {65}},\ \bibinfo
  {pages} {043514} (\bibinfo {year} {2002})},\ \Eprint
  {https://arxiv.org/abs/hep-ph/0003298v1} {arXiv:hep-ph/0003298v1 [hep-ph]}
  \BibitemShut {NoStop}%
\bibitem [{\citenamefont {Avelino}\ \emph {et~al.}(2005)\citenamefont
  {Avelino}, \citenamefont {Martins},\ and\ \citenamefont
  {Oliveira}}]{AvelinoMartins}%
  \BibitemOpen
  \bibfield  {author} {\bibinfo {author} {\bibfnamefont {P.~P.}\ \bibnamefont
  {Avelino}}, \bibinfo {author} {\bibfnamefont {C.~J. A.~P.}\ \bibnamefont
  {Martins}},\ and\ \bibinfo {author} {\bibfnamefont {J.~C. R.~E.}\
  \bibnamefont {Oliveira}},\ }\bibfield  {title} {\bibinfo {title} {One-scale
  model for domain wall network evolution},\ }\href
  {https://doi.org/10.1103/PhysRevD.72.083506} {\bibfield  {journal} {\bibinfo
  {journal} {Phys. Rev. D}\ }\textbf {\bibinfo {volume} {72}},\ \bibinfo
  {pages} {083506} (\bibinfo {year} {2005})},\ \Eprint
  {https://arxiv.org/abs/hep-ph/0507272v1} {arXiv:hep-ph/0507272v1 [hep-ph]}
  \BibitemShut {NoStop}%
\bibitem [{\citenamefont {Avelino}\ and\ \citenamefont
  {Sousa}(2011)}]{Avelino:2011ev}%
  \BibitemOpen
  \bibfield  {author} {\bibinfo {author} {\bibfnamefont {P.~P.}\ \bibnamefont
  {Avelino}}\ and\ \bibinfo {author} {\bibfnamefont {L.}~\bibnamefont
  {Sousa}},\ }\bibfield  {title} {\bibinfo {title} {{Domain wall network
  evolution in (N+1)-dimensional FRW universes}},\ }\href
  {https://doi.org/10.1103/PhysRevD.83.043530} {\bibfield  {journal} {\bibinfo
  {journal} {Phys. Rev. D}\ }\textbf {\bibinfo {volume} {83}},\ \bibinfo
  {pages} {043530} (\bibinfo {year} {2011})},\ \Eprint
  {https://arxiv.org/abs/1101.3360} {arXiv:1101.3360 [hep-th]} \BibitemShut
  {NoStop}%
\bibitem [{\citenamefont {Sousa}\ and\ \citenamefont
  {Avelino}(2011{\natexlab{a}})}]{Sousa:2011ew}%
  \BibitemOpen
  \bibfield  {author} {\bibinfo {author} {\bibfnamefont {L.}~\bibnamefont
  {Sousa}}\ and\ \bibinfo {author} {\bibfnamefont {P.~P.}\ \bibnamefont
  {Avelino}},\ }\bibfield  {title} {\bibinfo {title} {{p-brane dynamics in
  (N+1)-dimensional FRW universes: a unified framework}},\ }\href
  {https://doi.org/10.1103/PhysRevD.83.103507} {\bibfield  {journal} {\bibinfo
  {journal} {Phys. Rev. D}\ }\textbf {\bibinfo {volume} {83}},\ \bibinfo
  {pages} {103507} (\bibinfo {year} {2011}{\natexlab{a}})},\ \Eprint
  {https://arxiv.org/abs/1103.1381} {arXiv:1103.1381 [hep-th]} \BibitemShut
  {NoStop}%
\bibitem [{\citenamefont {Sousa}\ and\ \citenamefont
  {Avelino}(2011{\natexlab{b}})}]{SousaAvelino2}%
  \BibitemOpen
  \bibfield  {author} {\bibinfo {author} {\bibfnamefont {L.}~\bibnamefont
  {Sousa}}\ and\ \bibinfo {author} {\bibfnamefont {P.~P.}\ \bibnamefont
  {Avelino}},\ }\bibfield  {title} {\bibinfo {title} {Cosmological evolution of
  $p$-brane networks},\ }\href {https://doi.org/10.1103/PhysRevD.84.063502}
  {\bibfield  {journal} {\bibinfo  {journal} {Phys. Rev. D}\ }\textbf {\bibinfo
  {volume} {84}},\ \bibinfo {pages} {063502} (\bibinfo {year}
  {2011}{\natexlab{b}})},\ \Eprint {https://arxiv.org/abs/1107.4582v1}
  {arXiv:1107.4582v1 [hep-th]} \BibitemShut {NoStop}%
\bibitem [{\citenamefont {Sousa}\ and\ \citenamefont
  {Avelino}(2017)}]{SousaAvelino3}%
  \BibitemOpen
  \bibfield  {author} {\bibinfo {author} {\bibfnamefont {L.}~\bibnamefont
  {Sousa}}\ and\ \bibinfo {author} {\bibfnamefont {P.~P.}\ \bibnamefont
  {Avelino}},\ }\bibfield  {title} {\bibinfo {title} {Revisiting the
  velocity-dependent one-scale model for monopoles},\ }\href
  {https://doi.org/10.1103/PhysRevD.96.023521} {\bibfield  {journal} {\bibinfo
  {journal} {Phys. Rev. D}\ }\textbf {\bibinfo {volume} {96}},\ \bibinfo
  {pages} {023521} (\bibinfo {year} {2017})},\ \Eprint
  {https://arxiv.org/abs/1703.09054v2} {arXiv:1703.09054v2 [astro-ph.CO]}
  \BibitemShut {NoStop}%
\bibitem [{\citenamefont {Correia}\ and\ \citenamefont
  {Martins}(2019)}]{CorreiaMartins}%
  \BibitemOpen
  \bibfield  {author} {\bibinfo {author} {\bibfnamefont {J.~R. C. C.~C.}\
  \bibnamefont {Correia}}\ and\ \bibinfo {author} {\bibfnamefont {C.~J. A.~P.}\
  \bibnamefont {Martins}},\ }\bibfield  {title} {\bibinfo {title} {Extending
  and calibrating the velocity dependent one-scale model for cosmic strings
  with one thousand field theory simulations},\ }\href
  {https://doi.org/10.1103/PhysRevD.100.103517} {\bibfield  {journal} {\bibinfo
   {journal} {Phys. Rev. D}\ }\textbf {\bibinfo {volume} {100}},\ \bibinfo
  {pages} {103517} (\bibinfo {year} {2019})},\ \Eprint
  {https://arxiv.org/abs/1911.03163} {arXiv:1911.03163 [astro-ph.CO]}
  \BibitemShut {NoStop}%
\bibitem [{\citenamefont {Kibble}(1985)}]{Kibble:1984hp}%
  \BibitemOpen
  \bibfield  {author} {\bibinfo {author} {\bibfnamefont {T.~W.~B.}\
  \bibnamefont {Kibble}},\ }\bibfield  {title} {\bibinfo {title} {{Evolution of
  a system of cosmic strings}},\ }\href
  {https://doi.org/10.1016/0550-3213(85)90596-6} {\bibfield  {journal}
  {\bibinfo  {journal} {Nucl. Phys. B}\ }\textbf {\bibinfo {volume} {252}},\
  \bibinfo {pages} {227} (\bibinfo {year} {1985})},\ \bibinfo {note} {[Erratum:
  Nucl.Phys.B 261, 750 (1985)]}\BibitemShut {NoStop}%
\bibitem [{\citenamefont {Caldwell}\ and\ \citenamefont
  {Allen}(1992)}]{Caldwell:1991jj}%
  \BibitemOpen
  \bibfield  {author} {\bibinfo {author} {\bibfnamefont {R.~R.}\ \bibnamefont
  {Caldwell}}\ and\ \bibinfo {author} {\bibfnamefont {B.}~\bibnamefont
  {Allen}},\ }\bibfield  {title} {\bibinfo {title} {{Cosmological constraints
  on cosmic string gravitational radiation}},\ }\href
  {https://doi.org/10.1103/PhysRevD.45.3447} {\bibfield  {journal} {\bibinfo
  {journal} {Phys. Rev. D}\ }\textbf {\bibinfo {volume} {45}},\ \bibinfo
  {pages} {3447} (\bibinfo {year} {1992})}\BibitemShut {NoStop}%
\bibitem [{\citenamefont {DePies}\ and\ \citenamefont
  {Hogan}(2007)}]{DePies:2007bm}%
  \BibitemOpen
  \bibfield  {author} {\bibinfo {author} {\bibfnamefont {M.~R.}\ \bibnamefont
  {DePies}}\ and\ \bibinfo {author} {\bibfnamefont {C.~J.}\ \bibnamefont
  {Hogan}},\ }\bibfield  {title} {\bibinfo {title} {{Stochastic Gravitational
  Wave Background from Light Cosmic Strings}},\ }\href
  {https://doi.org/10.1103/PhysRevD.75.125006} {\bibfield  {journal} {\bibinfo
  {journal} {Phys. Rev. D}\ }\textbf {\bibinfo {volume} {75}},\ \bibinfo
  {pages} {125006} (\bibinfo {year} {2007})},\ \Eprint
  {https://arxiv.org/abs/astro-ph/0702335} {arXiv:astro-ph/0702335}
  \BibitemShut {NoStop}%
\bibitem [{\citenamefont {Sanidas}\ \emph {et~al.}(2012)\citenamefont
  {Sanidas}, \citenamefont {Battye},\ and\ \citenamefont
  {Stappers}}]{Sanidas:2012ee}%
  \BibitemOpen
  \bibfield  {author} {\bibinfo {author} {\bibfnamefont {S.~A.}\ \bibnamefont
  {Sanidas}}, \bibinfo {author} {\bibfnamefont {R.~A.}\ \bibnamefont
  {Battye}},\ and\ \bibinfo {author} {\bibfnamefont {B.~W.}\ \bibnamefont
  {Stappers}},\ }\bibfield  {title} {\bibinfo {title} {{Constraints on cosmic
  string tension imposed by the limit on the stochastic gravitational wave
  background from the European Pulsar Timing Array}},\ }\href
  {https://doi.org/10.1103/PhysRevD.85.122003} {\bibfield  {journal} {\bibinfo
  {journal} {Phys. Rev. D}\ }\textbf {\bibinfo {volume} {85}},\ \bibinfo
  {pages} {122003} (\bibinfo {year} {2012})},\ \Eprint
  {https://arxiv.org/abs/1201.2419} {arXiv:1201.2419 [astro-ph.CO]}
  \BibitemShut {NoStop}%
\bibitem [{\citenamefont {Kuroyanagi}\ \emph {et~al.}(2012)\citenamefont
  {Kuroyanagi}, \citenamefont {Miyamoto}, \citenamefont {Sekiguchi},
  \citenamefont {Takahashi},\ and\ \citenamefont {Silk}}]{Kuroyanagi:2012wm}%
  \BibitemOpen
  \bibfield  {author} {\bibinfo {author} {\bibfnamefont {S.}~\bibnamefont
  {Kuroyanagi}}, \bibinfo {author} {\bibfnamefont {K.}~\bibnamefont
  {Miyamoto}}, \bibinfo {author} {\bibfnamefont {T.}~\bibnamefont {Sekiguchi}},
  \bibinfo {author} {\bibfnamefont {K.}~\bibnamefont {Takahashi}},\ and\
  \bibinfo {author} {\bibfnamefont {J.}~\bibnamefont {Silk}},\ }\bibfield
  {title} {\bibinfo {title} {{Forecast constraints on cosmic string parameters
  from gravitational wave direct detection experiments}},\ }\href
  {https://doi.org/10.1103/PhysRevD.86.023503} {\bibfield  {journal} {\bibinfo
  {journal} {Phys. Rev. D}\ }\textbf {\bibinfo {volume} {86}},\ \bibinfo
  {pages} {023503} (\bibinfo {year} {2012})},\ \Eprint
  {https://arxiv.org/abs/1202.3032} {arXiv:1202.3032 [astro-ph.CO]}
  \BibitemShut {NoStop}%
\bibitem [{\citenamefont {Sousa}\ and\ \citenamefont
  {Avelino}(2013)}]{Sousa:2013aaa}%
  \BibitemOpen
  \bibfield  {author} {\bibinfo {author} {\bibfnamefont {L.}~\bibnamefont
  {Sousa}}\ and\ \bibinfo {author} {\bibfnamefont {P.~P.}\ \bibnamefont
  {Avelino}},\ }\bibfield  {title} {\bibinfo {title} {{Stochastic Gravitational
  Wave Background generated by Cosmic String Networks: Velocity-Dependent
  One-Scale model versus Scale-Invariant Evolution}},\ }\href
  {https://doi.org/10.1103/PhysRevD.88.023516} {\bibfield  {journal} {\bibinfo
  {journal} {Phys. Rev. D}\ }\textbf {\bibinfo {volume} {88}},\ \bibinfo
  {pages} {023516} (\bibinfo {year} {2013})},\ \Eprint
  {https://arxiv.org/abs/1304.2445} {arXiv:1304.2445 [astro-ph.CO]}
  \BibitemShut {NoStop}%
\bibitem [{\citenamefont {Blanco-Pillado}\ \emph {et~al.}(2014)\citenamefont
  {Blanco-Pillado}, \citenamefont {Olum},\ and\ \citenamefont
  {Shlaer}}]{Blanco-Pillado:2013qja}%
  \BibitemOpen
  \bibfield  {author} {\bibinfo {author} {\bibfnamefont {J.~J.}\ \bibnamefont
  {Blanco-Pillado}}, \bibinfo {author} {\bibfnamefont {K.~D.}\ \bibnamefont
  {Olum}},\ and\ \bibinfo {author} {\bibfnamefont {B.}~\bibnamefont {Shlaer}},\
  }\bibfield  {title} {\bibinfo {title} {{The number of cosmic string loops}},\
  }\href {https://doi.org/10.1103/PhysRevD.89.023512} {\bibfield  {journal}
  {\bibinfo  {journal} {Phys. Rev. D}\ }\textbf {\bibinfo {volume} {89}},\
  \bibinfo {pages} {023512} (\bibinfo {year} {2014})},\ \Eprint
  {https://arxiv.org/abs/1309.6637} {arXiv:1309.6637 [astro-ph.CO]}
  \BibitemShut {NoStop}%
\bibitem [{\citenamefont {Sousa}\ and\ \citenamefont
  {Avelino}(2014)}]{Sousa:2014gka}%
  \BibitemOpen
  \bibfield  {author} {\bibinfo {author} {\bibfnamefont {L.}~\bibnamefont
  {Sousa}}\ and\ \bibinfo {author} {\bibfnamefont {P.~P.}\ \bibnamefont
  {Avelino}},\ }\bibfield  {title} {\bibinfo {title} {{Stochastic gravitational
  wave background generated by cosmic string networks: The small-loop
  regime}},\ }\href {https://doi.org/10.1103/PhysRevD.89.083503} {\bibfield
  {journal} {\bibinfo  {journal} {Phys. Rev. D}\ }\textbf {\bibinfo {volume}
  {89}},\ \bibinfo {pages} {083503} (\bibinfo {year} {2014})},\ \Eprint
  {https://arxiv.org/abs/1403.2621} {arXiv:1403.2621 [astro-ph.CO]}
  \BibitemShut {NoStop}%
\bibitem [{\citenamefont {Sousa}\ \emph {et~al.}(2020)\citenamefont {Sousa},
  \citenamefont {Avelino},\ and\ \citenamefont {Guedes}}]{Sousa:2020sxs}%
  \BibitemOpen
  \bibfield  {author} {\bibinfo {author} {\bibfnamefont {L.}~\bibnamefont
  {Sousa}}, \bibinfo {author} {\bibfnamefont {P.~P.}\ \bibnamefont {Avelino}},\
  and\ \bibinfo {author} {\bibfnamefont {G.~S.~F.}\ \bibnamefont {Guedes}},\
  }\bibfield  {title} {\bibinfo {title} {{Full analytical approximation to the
  stochastic gravitational wave background generated by cosmic string
  networks}},\ }\href {https://doi.org/10.1103/PhysRevD.101.103508} {\bibfield
  {journal} {\bibinfo  {journal} {Phys. Rev. D}\ }\textbf {\bibinfo {volume}
  {101}},\ \bibinfo {pages} {103508} (\bibinfo {year} {2020})},\ \Eprint
  {https://arxiv.org/abs/2002.01079} {arXiv:2002.01079 [astro-ph.CO]}
  \BibitemShut {NoStop}%
\bibitem [{\citenamefont {Lorenz}\ \emph {et~al.}(2010)\citenamefont {Lorenz},
  \citenamefont {Ringeval},\ and\ \citenamefont
  {Sakellariadou}}]{Lorenz:2010sm}%
  \BibitemOpen
  \bibfield  {author} {\bibinfo {author} {\bibfnamefont {L.}~\bibnamefont
  {Lorenz}}, \bibinfo {author} {\bibfnamefont {C.}~\bibnamefont {Ringeval}},\
  and\ \bibinfo {author} {\bibfnamefont {M.}~\bibnamefont {Sakellariadou}},\
  }\bibfield  {title} {\bibinfo {title} {{Cosmic string loop distribution on
  all length scales and at any redshift}},\ }\href
  {https://doi.org/10.1088/1475-7516/2010/10/003} {\bibfield  {journal}
  {\bibinfo  {journal} {JCAP}\ }\textbf {\bibinfo {volume} {10}},\ \bibinfo
  {pages} {003}},\ \Eprint {https://arxiv.org/abs/1006.0931} {arXiv:1006.0931
  [astro-ph.CO]} \BibitemShut {NoStop}%
\bibitem [{\citenamefont {Blanco-Pillado}\ and\ \citenamefont
  {Olum}(2020)}]{Blanco-Pillado:2019tbi}%
  \BibitemOpen
  \bibfield  {author} {\bibinfo {author} {\bibfnamefont {J.~J.}\ \bibnamefont
  {Blanco-Pillado}}\ and\ \bibinfo {author} {\bibfnamefont {K.~D.}\
  \bibnamefont {Olum}},\ }\bibfield  {title} {\bibinfo {title} {{Direct
  determination of cosmic string loop density from simulations}},\ }\href
  {https://doi.org/10.1103/PhysRevD.101.103018} {\bibfield  {journal} {\bibinfo
   {journal} {Phys. Rev. D}\ }\textbf {\bibinfo {volume} {101}},\ \bibinfo
  {pages} {103018} (\bibinfo {year} {2020})},\ \Eprint
  {https://arxiv.org/abs/1912.10017} {arXiv:1912.10017 [astro-ph.CO]}
  \BibitemShut {NoStop}%
\bibitem [{\citenamefont {Auclair}\ \emph {et~al.}(2020)\citenamefont {Auclair}
  \emph {et~al.}}]{Auclair:2019wcv}%
  \BibitemOpen
  \bibfield  {author} {\bibinfo {author} {\bibfnamefont {P.}~\bibnamefont
  {Auclair}} \emph {et~al.},\ }\bibfield  {title} {\bibinfo {title} {{Probing
  the gravitational wave background from cosmic strings with LISA}},\ }\href
  {https://doi.org/10.1088/1475-7516/2020/04/034} {\bibfield  {journal}
  {\bibinfo  {journal} {JCAP}\ }\textbf {\bibinfo {volume} {04}},\ \bibinfo
  {pages} {034}},\ \Eprint {https://arxiv.org/abs/1909.00819} {arXiv:1909.00819
  [astro-ph.CO]} \BibitemShut {NoStop}%
\bibitem [{\citenamefont {Blanco-Pillado}\ \emph {et~al.}(2019)\citenamefont
  {Blanco-Pillado}, \citenamefont {Olum},\ and\ \citenamefont
  {Wachter}}]{Blanco-Pillado:2019vcs}%
  \BibitemOpen
  \bibfield  {author} {\bibinfo {author} {\bibfnamefont {J.~J.}\ \bibnamefont
  {Blanco-Pillado}}, \bibinfo {author} {\bibfnamefont {K.~D.}\ \bibnamefont
  {Olum}},\ and\ \bibinfo {author} {\bibfnamefont {J.~M.}\ \bibnamefont
  {Wachter}},\ }\bibfield  {title} {\bibinfo {title} {{Energy-conservation
  constraints on cosmic string loop production and distribution functions}},\
  }\href {https://doi.org/10.1103/PhysRevD.100.123526} {\bibfield  {journal}
  {\bibinfo  {journal} {Phys. Rev. D}\ }\textbf {\bibinfo {volume} {100}},\
  \bibinfo {pages} {123526} (\bibinfo {year} {2019})},\ \Eprint
  {https://arxiv.org/abs/1907.09373} {arXiv:1907.09373 [astro-ph.CO]}
  \BibitemShut {NoStop}%
\bibitem [{\citenamefont {Hindmarsh}\ \emph {et~al.}(2009)\citenamefont
  {Hindmarsh}, \citenamefont {Stuckey},\ and\ \citenamefont
  {Bevis}}]{HindmarshStuckeyBevis}%
  \BibitemOpen
  \bibfield  {author} {\bibinfo {author} {\bibfnamefont {M.}~\bibnamefont
  {Hindmarsh}}, \bibinfo {author} {\bibfnamefont {S.}~\bibnamefont {Stuckey}},\
  and\ \bibinfo {author} {\bibfnamefont {N.}~\bibnamefont {Bevis}},\ }\bibfield
   {title} {\bibinfo {title} {Abelian higgs cosmic strings: Small-scale
  structure and loops},\ }\href {https://doi.org/10.1103/PhysRevD.79.123504}
  {\bibfield  {journal} {\bibinfo  {journal} {Phys. Rev. D}\ }\textbf {\bibinfo
  {volume} {79}},\ \bibinfo {pages} {123504} (\bibinfo {year} {2009})},\
  \Eprint {https://arxiv.org/abs/0812.1929} {arXiv:0812.1929 [hep-th]}
  \BibitemShut {NoStop}%
\bibitem [{\citenamefont {Quashnock}\ and\ \citenamefont
  {Spergel}(1990)}]{QuashnockSpergel}%
  \BibitemOpen
  \bibfield  {author} {\bibinfo {author} {\bibfnamefont {J.~M.}\ \bibnamefont
  {Quashnock}}\ and\ \bibinfo {author} {\bibfnamefont {D.~N.}\ \bibnamefont
  {Spergel}},\ }\bibfield  {title} {\bibinfo {title} {Gravitational
  self-interactions of cosmic strings},\ }\href
  {https://doi.org/10.1103/PhysRevD.42.2505} {\bibfield  {journal} {\bibinfo
  {journal} {Phys. Rev. D}\ }\textbf {\bibinfo {volume} {42}},\ \bibinfo
  {pages} {2505} (\bibinfo {year} {1990})}\BibitemShut {NoStop}%
\bibitem [{\citenamefont {Scherrer}\ \emph {et~al.}(1990)\citenamefont
  {Scherrer}, \citenamefont {Quashnock}, \citenamefont {Spergel},\ and\
  \citenamefont {Press}}]{ScherrerQuashnockSpergel}%
  \BibitemOpen
  \bibfield  {author} {\bibinfo {author} {\bibfnamefont {R.~J.}\ \bibnamefont
  {Scherrer}}, \bibinfo {author} {\bibfnamefont {J.~M.}\ \bibnamefont
  {Quashnock}}, \bibinfo {author} {\bibfnamefont {D.~N.}\ \bibnamefont
  {Spergel}},\ and\ \bibinfo {author} {\bibfnamefont {W.~H.}\ \bibnamefont
  {Press}},\ }\bibfield  {title} {\bibinfo {title} {Properties of realistic
  cosmic-string loops},\ }\href {https://doi.org/10.1103/PhysRevD.42.1908}
  {\bibfield  {journal} {\bibinfo  {journal} {Phys. Rev. D}\ }\textbf {\bibinfo
  {volume} {42}},\ \bibinfo {pages} {1908} (\bibinfo {year}
  {1990})}\BibitemShut {NoStop}%
\bibitem [{\citenamefont {{Planck Collaboration}}\ \emph
  {et~al.}(2020)\citenamefont {{Planck Collaboration}}, \citenamefont
  {{Aghanim, N.}} \emph {et~al.}}]{Planck2020}%
  \BibitemOpen
  \bibfield  {author} {\bibinfo {author} {\bibnamefont {{Planck
  Collaboration}}}, \bibinfo {author} {\bibnamefont {{Aghanim, N.}}}, \emph
  {et~al.},\ }\bibfield  {title} {\bibinfo {title} {Planck 2018 results - vi.
  cosmological parameters},\ }\href
  {https://doi.org/10.1051/0004-6361/201833910} {\bibfield  {journal} {\bibinfo
   {journal} {A\&A}\ }\textbf {\bibinfo {volume} {641}},\ \bibinfo {pages} {A6}
  (\bibinfo {year} {2020})},\ \Eprint {https://arxiv.org/abs/1807.06209}
  {arXiv:1807.06209 [astro-ph.CO]} \BibitemShut {NoStop}%
\bibitem [{\citenamefont {Pogosian}\ \emph {et~al.}(2009)\citenamefont
  {Pogosian}, \citenamefont {Tye}, \citenamefont {Wasserman},\ and\
  \citenamefont {Wyman}}]{PogosianTyeWassermanWyman2}%
  \BibitemOpen
  \bibfield  {author} {\bibinfo {author} {\bibfnamefont {L.}~\bibnamefont
  {Pogosian}}, \bibinfo {author} {\bibfnamefont {S.-H.~H.}\ \bibnamefont
  {Tye}}, \bibinfo {author} {\bibfnamefont {I.}~\bibnamefont {Wasserman}},\
  and\ \bibinfo {author} {\bibfnamefont {M.}~\bibnamefont {Wyman}},\ }\bibfield
   {title} {\bibinfo {title} {Cosmic strings as the source of small-scale
  microwave background anisotropy},\ }\href
  {https://doi.org/10.1088/1475-7516/2009/02/013} {\bibfield  {journal}
  {\bibinfo  {journal} {Journal of Cosmology and Astroparticle Physics}\
  }\textbf {\bibinfo {volume} {2009}}\bibfield  {number} {\bibinfo  {number} {
  (02)},\ \bibinfo {pages} {013}},\ }\Eprint {https://arxiv.org/abs/0804.0810}
  {arXiv:0804.0810 [astro-ph]} \BibitemShut {NoStop}%
\bibitem [{\citenamefont {{Planck Collaboration}}\ \emph
  {et~al.}(2014)\citenamefont {{Planck Collaboration}}, \citenamefont {{Ade, P.
  A. R.}} \emph {et~al.}}]{Ade:2013xla}%
  \BibitemOpen
  \bibfield  {author} {\bibinfo {author} {\bibnamefont {{Planck
  Collaboration}}}, \bibinfo {author} {\bibnamefont {{Ade, P. A. R.}}}, \emph
  {et~al.},\ }\bibfield  {title} {\bibinfo {title} {Planck 2013 results. xxv.
  searches for cosmic strings and other topological defects},\ }\href
  {https://doi.org/10.1051/0004-6361/201321621} {\bibfield  {journal} {\bibinfo
   {journal} {A\&A}\ }\textbf {\bibinfo {volume} {571}},\ \bibinfo {pages}
  {A25} (\bibinfo {year} {2014})},\ \Eprint {https://arxiv.org/abs/1303.5085}
  {arXiv:1303.5085 [astro-ph.CO]} \BibitemShut {NoStop}%
\bibitem [{\citenamefont {Dvali}\ and\ \citenamefont {Tye}(1999)}]{DvaliTye}%
  \BibitemOpen
  \bibfield  {author} {\bibinfo {author} {\bibfnamefont {G.}~\bibnamefont
  {Dvali}}\ and\ \bibinfo {author} {\bibfnamefont {S.-H.}\ \bibnamefont
  {Tye}},\ }\bibfield  {title} {\bibinfo {title} {Brane inflation},\ }\href
  {https://doi.org/https://doi.org/10.1016/S0370-2693(99)00132-X} {\bibfield
  {journal} {\bibinfo  {journal} {Physics Letters B}\ }\textbf {\bibinfo
  {volume} {450}},\ \bibinfo {pages} {72} (\bibinfo {year} {1999})},\ \Eprint
  {https://arxiv.org/abs/hep-ph/9812483} {arXiv:hep-ph/9812483 [hep-ph]}
  \BibitemShut {NoStop}%
\bibitem [{\citenamefont {Burgess}\ \emph {et~al.}(2001)\citenamefont
  {Burgess}, \citenamefont {Majumdar}, \citenamefont {Nolte}, \citenamefont
  {Quevedo}, \citenamefont {Rajesh},\ and\ \citenamefont
  {Zhang}}]{BurgessMajumdarNolteQuevedoRajeshZhang}%
  \BibitemOpen
  \bibfield  {author} {\bibinfo {author} {\bibfnamefont {C.~P.}\ \bibnamefont
  {Burgess}}, \bibinfo {author} {\bibfnamefont {M.}~\bibnamefont {Majumdar}},
  \bibinfo {author} {\bibfnamefont {D.}~\bibnamefont {Nolte}}, \bibinfo
  {author} {\bibfnamefont {F.}~\bibnamefont {Quevedo}}, \bibinfo {author}
  {\bibfnamefont {G.}~\bibnamefont {Rajesh}},\ and\ \bibinfo {author}
  {\bibfnamefont {R.-J.}\ \bibnamefont {Zhang}},\ }\bibfield  {title} {\bibinfo
  {title} {The inflationary brane-antibrane universe},\ }\href
  {https://doi.org/10.1088/1126-6708/2001/07/047} {\bibfield  {journal}
  {\bibinfo  {journal} {Journal of High Energy Physics}\ }\textbf {\bibinfo
  {volume} {2001}},\ \bibinfo {pages} {047} (\bibinfo {year} {2001})},\ \Eprint
  {https://arxiv.org/abs/hep-th/0105204} {arXiv:hep-th/0105204 [hep-th]}
  \BibitemShut {NoStop}%
\bibitem [{\citenamefont {Sarangi}\ and\ \citenamefont
  {Tye}(2002)}]{SarangiTye}%
  \BibitemOpen
  \bibfield  {author} {\bibinfo {author} {\bibfnamefont {S.}~\bibnamefont
  {Sarangi}}\ and\ \bibinfo {author} {\bibfnamefont {S.-H.}\ \bibnamefont
  {Tye}},\ }\bibfield  {title} {\bibinfo {title} {Cosmic string production
  towards the end of brane inflation},\ }\href
  {https://doi.org/https://doi.org/10.1016/S0370-2693(02)01824-5} {\bibfield
  {journal} {\bibinfo  {journal} {Physics Letters B}\ }\textbf {\bibinfo
  {volume} {536}},\ \bibinfo {pages} {185} (\bibinfo {year} {2002})},\ \Eprint
  {https://arxiv.org/abs/hep-th/0204074} {arXiv:hep-th/0204074 [hep-th]}
  \BibitemShut {NoStop}%
\bibitem [{\citenamefont {Jones}\ \emph {et~al.}(2003)\citenamefont {Jones},
  \citenamefont {Stoica},\ and\ \citenamefont {Tye}}]{JonesStoicaTye}%
  \BibitemOpen
  \bibfield  {author} {\bibinfo {author} {\bibfnamefont {N.~T.}\ \bibnamefont
  {Jones}}, \bibinfo {author} {\bibfnamefont {H.}~\bibnamefont {Stoica}},\ and\
  \bibinfo {author} {\bibfnamefont {S.-H.}\ \bibnamefont {Tye}},\ }\bibfield
  {title} {\bibinfo {title} {The production, spectrum and evolution of cosmic
  strings in brane inflation},\ }\href
  {https://doi.org/https://doi.org/10.1016/S0370-2693(03)00592-6} {\bibfield
  {journal} {\bibinfo  {journal} {Physics Letters B}\ }\textbf {\bibinfo
  {volume} {563}},\ \bibinfo {pages} {6} (\bibinfo {year} {2003})},\ \Eprint
  {https://arxiv.org/abs/hep-th/0303269} {arXiv:hep-th/0303269 [hep-th]}
  \BibitemShut {NoStop}%
\bibitem [{\citenamefont {Dvali}\ and\ \citenamefont
  {Vilenkin}(2004)}]{DvaliVilenkin}%
  \BibitemOpen
  \bibfield  {author} {\bibinfo {author} {\bibfnamefont {G.}~\bibnamefont
  {Dvali}}\ and\ \bibinfo {author} {\bibfnamefont {A.}~\bibnamefont
  {Vilenkin}},\ }\bibfield  {title} {\bibinfo {title} {Formation and evolution
  of {cosmicDstrings}},\ }\href {https://doi.org/10.1088/1475-7516/2004/03/010}
  {\bibfield  {journal} {\bibinfo  {journal} {Journal of Cosmology and
  Astroparticle Physics}\ }\textbf {\bibinfo {volume} {2004}}\bibfield
  {number} {\bibinfo  {number} { (03)},\ \bibinfo {pages} {010}},\ }\Eprint
  {https://arxiv.org/abs/hep-th/0312007} {arXiv:hep-th/0312007 [hep-th]}
  \BibitemShut {NoStop}%
\bibitem [{\citenamefont {Dvali}\ \emph {et~al.}(2004)\citenamefont {Dvali},
  \citenamefont {Kallosh},\ and\ \citenamefont
  {Proeyen}}]{DvaliKalloshProeyen}%
  \BibitemOpen
  \bibfield  {author} {\bibinfo {author} {\bibfnamefont {G.}~\bibnamefont
  {Dvali}}, \bibinfo {author} {\bibfnamefont {R.}~\bibnamefont {Kallosh}},\
  and\ \bibinfo {author} {\bibfnamefont {A.~V.}\ \bibnamefont {Proeyen}},\
  }\bibfield  {title} {\bibinfo {title} {D-term strings},\ }\href
  {https://doi.org/10.1088/1126-6708/2004/01/035} {\bibfield  {journal}
  {\bibinfo  {journal} {Journal of High Energy Physics}\ }\textbf {\bibinfo
  {volume} {2004}},\ \bibinfo {pages} {035} (\bibinfo {year} {2004})},\ \Eprint
  {https://arxiv.org/abs/hep-th/0312005} {arXiv:hep-th/0312005 [hep-th]}
  \BibitemShut {NoStop}%
\bibitem [{\citenamefont {Jackson}\ \emph {et~al.}(2005)\citenamefont
  {Jackson}, \citenamefont {Jones},\ and\ \citenamefont
  {Polchinski}}]{JacksonJonesPolchinski}%
  \BibitemOpen
  \bibfield  {author} {\bibinfo {author} {\bibfnamefont {M.~G.}\ \bibnamefont
  {Jackson}}, \bibinfo {author} {\bibfnamefont {N.~T.}\ \bibnamefont {Jones}},\
  and\ \bibinfo {author} {\bibfnamefont {J.}~\bibnamefont {Polchinski}},\
  }\bibfield  {title} {\bibinfo {title} {Collisions of cosmic f- and
  d-strings},\ }\href {https://doi.org/10.1088/1126-6708/2005/10/013}
  {\bibfield  {journal} {\bibinfo  {journal} {Journal of High Energy Physics}\
  }\textbf {\bibinfo {volume} {2005}},\ \bibinfo {pages} {013} (\bibinfo {year}
  {2005})},\ \Eprint {https://arxiv.org/abs/hep-th/0405229}
  {arXiv:hep-th/0405229 [hep-th]} \BibitemShut {NoStop}%
\bibitem [{\citenamefont {Hanany}\ and\ \citenamefont
  {Hashimoto}(2005)}]{HananyHashimoto}%
  \BibitemOpen
  \bibfield  {author} {\bibinfo {author} {\bibfnamefont {A.}~\bibnamefont
  {Hanany}}\ and\ \bibinfo {author} {\bibfnamefont {K.}~\bibnamefont
  {Hashimoto}},\ }\bibfield  {title} {\bibinfo {title} {Reconnection of
  colliding cosmic strings},\ }\href
  {https://doi.org/10.1088/1126-6708/2005/06/021} {\bibfield  {journal}
  {\bibinfo  {journal} {Journal of High Energy Physics}\ }\textbf {\bibinfo
  {volume} {2005}},\ \bibinfo {pages} {021} (\bibinfo {year} {2005})},\ \Eprint
  {https://arxiv.org/abs/hep-th/0501031} {arXiv:hep-th/0501031 [hep-th]}
  \BibitemShut {NoStop}%
\bibitem [{\citenamefont {Jackson}(2007)}]{Jackson_2007}%
  \BibitemOpen
  \bibfield  {author} {\bibinfo {author} {\bibfnamefont {M.~G.}\ \bibnamefont
  {Jackson}},\ }\bibfield  {title} {\bibinfo {title} {Interactions of cosmic
  superstrings},\ }\href {https://doi.org/10.1088/1126-6708/2007/09/035}
  {\bibfield  {journal} {\bibinfo  {journal} {Journal of High Energy Physics}\
  }\textbf {\bibinfo {volume} {2007}},\ \bibinfo {pages} {035} (\bibinfo {year}
  {2007})},\ \Eprint {https://arxiv.org/abs/0706.1264} {arXiv:0706.1264
  [hep-th]} \BibitemShut {NoStop}%
\bibitem [{\citenamefont {Avgoustidis}\ and\ \citenamefont
  {Shellard}(2005)}]{AvgoustidisShellard2}%
  \BibitemOpen
  \bibfield  {author} {\bibinfo {author} {\bibfnamefont {A.}~\bibnamefont
  {Avgoustidis}}\ and\ \bibinfo {author} {\bibfnamefont {E.~P.~S.}\
  \bibnamefont {Shellard}},\ }\bibfield  {title} {\bibinfo {title} {Cosmic
  string evolution in higher dimensions},\ }\href
  {https://doi.org/10.1103/PhysRevD.71.123513} {\bibfield  {journal} {\bibinfo
  {journal} {Phys. Rev. D}\ }\textbf {\bibinfo {volume} {71}},\ \bibinfo
  {pages} {123513} (\bibinfo {year} {2005})},\ \Eprint
  {https://arxiv.org/abs/hep-ph/0410349} {arXiv:hep-ph/0410349 [hep-ph]}
  \BibitemShut {NoStop}%
\bibitem [{\citenamefont {Tye}\ \emph {et~al.}(2005)\citenamefont {Tye},
  \citenamefont {Wasserman},\ and\ \citenamefont {Wyman}}]{TyeWassermanWyman}%
  \BibitemOpen
  \bibfield  {author} {\bibinfo {author} {\bibfnamefont {S.-H.~H.}\
  \bibnamefont {Tye}}, \bibinfo {author} {\bibfnamefont {I.}~\bibnamefont
  {Wasserman}},\ and\ \bibinfo {author} {\bibfnamefont {M.}~\bibnamefont
  {Wyman}},\ }\bibfield  {title} {\bibinfo {title} {Scaling of multitension
  cosmic superstring networks},\ }\href
  {https://doi.org/10.1103/PhysRevD.71.103508} {\bibfield  {journal} {\bibinfo
  {journal} {Phys. Rev. D}\ }\textbf {\bibinfo {volume} {71}},\ \bibinfo
  {pages} {103508} (\bibinfo {year} {2005})},\ \Eprint
  {https://arxiv.org/abs/astro-ph/0503506} {arXiv:astro-ph/0503506 [astro-ph]}
  \BibitemShut {NoStop}%
\bibitem [{\citenamefont {Avgoustidis}\ and\ \citenamefont
  {Shellard}(2008)}]{AvgoustidisShellard3}%
  \BibitemOpen
  \bibfield  {author} {\bibinfo {author} {\bibfnamefont {A.}~\bibnamefont
  {Avgoustidis}}\ and\ \bibinfo {author} {\bibfnamefont {E.~P.~S.}\
  \bibnamefont {Shellard}},\ }\bibfield  {title} {\bibinfo {title}
  {Velocity-dependent models for non-abelian/entangled string networks},\
  }\href {https://doi.org/10.1103/PhysRevD.78.103510} {\bibfield  {journal}
  {\bibinfo  {journal} {Phys. Rev. D}\ }\textbf {\bibinfo {volume} {78}},\
  \bibinfo {pages} {103510} (\bibinfo {year} {2008})},\ \Eprint
  {https://arxiv.org/abs/0705.3395} {arXiv:0705.3395 [astro-ph]} \BibitemShut
  {NoStop}%
\bibitem [{\citenamefont {Pourtsidou}\ \emph {et~al.}(2011)\citenamefont
  {Pourtsidou}, \citenamefont {Avgoustidis}, \citenamefont {Copeland},
  \citenamefont {Pogosian},\ and\ \citenamefont
  {Steer}}]{PourtsidouAvgoustidisCopelandPogosianSteer}%
  \BibitemOpen
  \bibfield  {author} {\bibinfo {author} {\bibfnamefont {A.}~\bibnamefont
  {Pourtsidou}}, \bibinfo {author} {\bibfnamefont {A.}~\bibnamefont
  {Avgoustidis}}, \bibinfo {author} {\bibfnamefont {E.~J.}\ \bibnamefont
  {Copeland}}, \bibinfo {author} {\bibfnamefont {L.}~\bibnamefont {Pogosian}},\
  and\ \bibinfo {author} {\bibfnamefont {D.~A.}\ \bibnamefont {Steer}},\
  }\bibfield  {title} {\bibinfo {title} {Scaling configurations of cosmic
  superstring networks and their cosmological implications},\ }\href
  {https://doi.org/10.1103/PhysRevD.83.063525} {\bibfield  {journal} {\bibinfo
  {journal} {Phys. Rev. D}\ }\textbf {\bibinfo {volume} {83}},\ \bibinfo
  {pages} {063525} (\bibinfo {year} {2011})},\ \Eprint
  {https://arxiv.org/abs/1012.5014} {arXiv:1012.5014 [astro-ph.CO]}
  \BibitemShut {NoStop}%
\bibitem [{\citenamefont {Rybak}\ \emph {et~al.}(2019)\citenamefont {Rybak},
  \citenamefont {Avgoustidis},\ and\ \citenamefont
  {Martins}}]{RybakAvgoustidisMartins2}%
  \BibitemOpen
  \bibfield  {author} {\bibinfo {author} {\bibfnamefont {I.~Y.}\ \bibnamefont
  {Rybak}}, \bibinfo {author} {\bibfnamefont {A.}~\bibnamefont {Avgoustidis}},\
  and\ \bibinfo {author} {\bibfnamefont {C.~J. A.~P.}\ \bibnamefont
  {Martins}},\ }\bibfield  {title} {\bibinfo {title} {Dynamics of junctions and
  the multitension velocity-dependent one-scale model},\ }\href
  {https://doi.org/10.1103/PhysRevD.99.063516} {\bibfield  {journal} {\bibinfo
  {journal} {Phys. Rev. D}\ }\textbf {\bibinfo {volume} {99}},\ \bibinfo
  {pages} {063516} (\bibinfo {year} {2019})},\ \Eprint
  {https://arxiv.org/abs/1812.04584} {arXiv:1812.04584 [astro-ph.CO]}
  \BibitemShut {NoStop}%
\bibitem [{\citenamefont {Sousa}\ and\ \citenamefont
  {Avelino}(2016)}]{Sousa:2016ggw}%
  \BibitemOpen
  \bibfield  {author} {\bibinfo {author} {\bibfnamefont {L.}~\bibnamefont
  {Sousa}}\ and\ \bibinfo {author} {\bibfnamefont {P.~P.}\ \bibnamefont
  {Avelino}},\ }\bibfield  {title} {\bibinfo {title} {{Probing Cosmic
  Superstrings with Gravitational Waves}},\ }\href
  {https://doi.org/10.1103/PhysRevD.94.063529} {\bibfield  {journal} {\bibinfo
  {journal} {Phys. Rev. D}\ }\textbf {\bibinfo {volume} {94}},\ \bibinfo
  {pages} {063529} (\bibinfo {year} {2016})},\ \Eprint
  {https://arxiv.org/abs/1606.05585} {arXiv:1606.05585 [astro-ph.CO]}
  \BibitemShut {NoStop}%
\bibitem [{\citenamefont {Avgoustidis}\ and\ \citenamefont
  {Shellard}(2006)}]{AvgoustidisShellard}%
  \BibitemOpen
  \bibfield  {author} {\bibinfo {author} {\bibfnamefont {A.}~\bibnamefont
  {Avgoustidis}}\ and\ \bibinfo {author} {\bibfnamefont {E.~P.~S.}\
  \bibnamefont {Shellard}},\ }\bibfield  {title} {\bibinfo {title} {Effect of
  reconnection probability on cosmic (super)string network density},\ }\href
  {https://doi.org/10.1103/PhysRevD.73.041301} {\bibfield  {journal} {\bibinfo
  {journal} {Phys. Rev. D}\ }\textbf {\bibinfo {volume} {73}},\ \bibinfo
  {pages} {041301(R)} (\bibinfo {year} {2006})},\ \Eprint
  {https://arxiv.org/abs/astro-ph/0512582} {arXiv:astro-ph/0512582 [astro-ph]}
  \BibitemShut {NoStop}%
\bibitem [{\citenamefont {Sakellariadou}(2005)}]{Sakellariadou_2005}%
  \BibitemOpen
  \bibfield  {author} {\bibinfo {author} {\bibfnamefont {M.}~\bibnamefont
  {Sakellariadou}},\ }\bibfield  {title} {\bibinfo {title} {A note on the
  evolution of cosmic string/superstring networks},\ }\href
  {https://doi.org/10.1088/1475-7516/2005/04/003} {\bibfield  {journal}
  {\bibinfo  {journal} {Journal of Cosmology and Astroparticle Physics}\
  }\textbf {\bibinfo {volume} {2005}}\bibfield  {number} {\bibinfo  {number} {
  (04)},\ \bibinfo {pages} {003}},\ }\Eprint
  {https://arxiv.org/abs/hep-th/0410234} {arXiv:hep-th/0410234 [hep-th]}
  \BibitemShut {NoStop}%
\bibitem [{\citenamefont {Avelino}\ and\ \citenamefont
  {Sousa}(2012)}]{Avelino:2012qy}%
  \BibitemOpen
  \bibfield  {author} {\bibinfo {author} {\bibfnamefont {P.~P.}\ \bibnamefont
  {Avelino}}\ and\ \bibinfo {author} {\bibfnamefont {L.}~\bibnamefont
  {Sousa}},\ }\bibfield  {title} {\bibinfo {title} {{Scaling laws for weakly
  interacting cosmic (super)string and p-brane networks}},\ }\href
  {https://doi.org/10.1103/PhysRevD.85.083525} {\bibfield  {journal} {\bibinfo
  {journal} {Phys. Rev. D}\ }\textbf {\bibinfo {volume} {85}},\ \bibinfo
  {pages} {083525} (\bibinfo {year} {2012})},\ \Eprint
  {https://arxiv.org/abs/1202.6298} {arXiv:1202.6298 [astro-ph.CO]}
  \BibitemShut {NoStop}%
\bibitem [{\citenamefont {Lentati}\ \emph {et~al.}(2015)\citenamefont {Lentati}
  \emph {et~al.}}]{Lentati:2015qwp}%
  \BibitemOpen
  \bibfield  {author} {\bibinfo {author} {\bibfnamefont {L.}~\bibnamefont
  {Lentati}} \emph {et~al.},\ }\bibfield  {title} {\bibinfo {title} {{European
  Pulsar Timing Array Limits On An Isotropic Stochastic Gravitational-Wave
  Background}},\ }\href {https://doi.org/10.1093/mnras/stv1538} {\bibfield
  {journal} {\bibinfo  {journal} {Mon. Not. Roy. Astron. Soc.}\ }\textbf
  {\bibinfo {volume} {453}},\ \bibinfo {pages} {2576} (\bibinfo {year}
  {2015})},\ \Eprint {https://arxiv.org/abs/1504.03692} {arXiv:1504.03692
  [astro-ph.CO]} \BibitemShut {NoStop}%
\bibitem [{\citenamefont {Arzoumanian}\ \emph {et~al.}(2016)\citenamefont
  {Arzoumanian} \emph {et~al.}}]{Arzoumanian:2015liz}%
  \BibitemOpen
  \bibfield  {author} {\bibinfo {author} {\bibfnamefont {Z.}~\bibnamefont
  {Arzoumanian}} \emph {et~al.} (\bibinfo {collaboration} {NANOGrav}),\
  }\bibfield  {title} {\bibinfo {title} {{The NANOGrav Nine-year Data Set:
  Limits on the Isotropic Stochastic Gravitational Wave Background}},\ }\href
  {https://doi.org/10.3847/0004-637X/821/1/13} {\bibfield  {journal} {\bibinfo
  {journal} {Astrophys. J.}\ }\textbf {\bibinfo {volume} {821}},\ \bibinfo
  {pages} {13} (\bibinfo {year} {2016})},\ \Eprint
  {https://arxiv.org/abs/1508.03024} {arXiv:1508.03024 [astro-ph.GA]}
  \BibitemShut {NoStop}%
\bibitem [{\citenamefont {Blanco-Pillado}\ \emph {et~al.}(2018)\citenamefont
  {Blanco-Pillado}, \citenamefont {Olum},\ and\ \citenamefont
  {Siemens}}]{Blanco-Pillado:2017rnf}%
  \BibitemOpen
  \bibfield  {author} {\bibinfo {author} {\bibfnamefont {J.~J.}\ \bibnamefont
  {Blanco-Pillado}}, \bibinfo {author} {\bibfnamefont {K.~D.}\ \bibnamefont
  {Olum}},\ and\ \bibinfo {author} {\bibfnamefont {X.}~\bibnamefont
  {Siemens}},\ }\bibfield  {title} {\bibinfo {title} {{New limits on cosmic
  strings from gravitational wave observation}},\ }\href
  {https://doi.org/10.1016/j.physletb.2018.01.050} {\bibfield  {journal}
  {\bibinfo  {journal} {Phys. Lett. B}\ }\textbf {\bibinfo {volume} {778}},\
  \bibinfo {pages} {392} (\bibinfo {year} {2018})},\ \Eprint
  {https://arxiv.org/abs/1709.02434} {arXiv:1709.02434 [astro-ph.CO]}
  \BibitemShut {NoStop}%
\bibitem [{\citenamefont {Fukuda}\ \emph {et~al.}(2020)\citenamefont {Fukuda},
  \citenamefont {Manohar}, \citenamefont {Murayama},\ and\ \citenamefont
  {Telem}}]{FukudaManoharMurayamaTelem}%
  \BibitemOpen
  \bibfield  {author} {\bibinfo {author} {\bibfnamefont {H.}~\bibnamefont
  {Fukuda}}, \bibinfo {author} {\bibfnamefont {A.~V.}\ \bibnamefont {Manohar}},
  \bibinfo {author} {\bibfnamefont {H.}~\bibnamefont {Murayama}},\ and\
  \bibinfo {author} {\bibfnamefont {O.}~\bibnamefont {Telem}},\ }\bibfield
  {title} {\bibinfo {title} {Axion strings are superconducting},\ }\href@noop
  {} {\  (\bibinfo {year} {2020})},\ \Eprint {https://arxiv.org/abs/2010.02763}
  {arXiv:2010.02763 [hep-ph]} \BibitemShut {NoStop}%
\bibitem [{\citenamefont {Abe}\ \emph {et~al.}(2020)\citenamefont {Abe},
  \citenamefont {Hamada},\ and\ \citenamefont {Yoshioka}}]{AbeHamadaYoshioka}%
  \BibitemOpen
  \bibfield  {author} {\bibinfo {author} {\bibfnamefont {Y.}~\bibnamefont
  {Abe}}, \bibinfo {author} {\bibfnamefont {Y.}~\bibnamefont {Hamada}},\ and\
  \bibinfo {author} {\bibfnamefont {K.}~\bibnamefont {Yoshioka}},\ }\bibfield
  {title} {\bibinfo {title} {Electroweak axion string and superconductivity},\
  }\href@noop {} {\  (\bibinfo {year} {2020})},\ \Eprint
  {https://arxiv.org/abs/2010.02834} {arXiv:2010.02834 [hep-ph]} \BibitemShut
  {NoStop}%
\bibitem [{\citenamefont {Davis}\ and\ \citenamefont
  {Shellard}(1989)}]{Davis:1988ij}%
  \BibitemOpen
  \bibfield  {author} {\bibinfo {author} {\bibfnamefont {R.~L.}\ \bibnamefont
  {Davis}}\ and\ \bibinfo {author} {\bibfnamefont {E.~P.~S.}\ \bibnamefont
  {Shellard}},\ }\bibfield  {title} {\bibinfo {title} {{Cosmic vortons}},\
  }\href {https://doi.org/10.1016/0550-3213(89)90594-4} {\bibfield  {journal}
  {\bibinfo  {journal} {Nucl. Phys.}\ }\textbf {\bibinfo {volume} {B323}},\
  \bibinfo {pages} {209} (\bibinfo {year} {1989})}\BibitemShut {NoStop}%
\bibitem [{\citenamefont {Brandenberger}\ \emph {et~al.}(1996)\citenamefont
  {Brandenberger}, \citenamefont {Carter}, \citenamefont {Davis},\ and\
  \citenamefont {Trodden}}]{Brandenberger:1996zp}%
  \BibitemOpen
  \bibfield  {author} {\bibinfo {author} {\bibfnamefont {R.~H.}\ \bibnamefont
  {Brandenberger}}, \bibinfo {author} {\bibfnamefont {B.}~\bibnamefont
  {Carter}}, \bibinfo {author} {\bibfnamefont {A.-C.}\ \bibnamefont {Davis}},\
  and\ \bibinfo {author} {\bibfnamefont {M.}~\bibnamefont {Trodden}},\
  }\bibfield  {title} {\bibinfo {title} {{Cosmic vortons and particle physics
  constraints}},\ }\href {https://doi.org/10.1103/PhysRevD.54.6059} {\bibfield
  {journal} {\bibinfo  {journal} {Phys. Rev.}\ }\textbf {\bibinfo {volume}
  {D54}},\ \bibinfo {pages} {6059} (\bibinfo {year} {1996})},\ \Eprint
  {https://arxiv.org/abs/hep-ph/9605382} {arXiv:hep-ph/9605382 [hep-ph]}
  \BibitemShut {NoStop}%
\bibitem [{\citenamefont {Auclair}\ \emph {et~al.}(2021)\citenamefont
  {Auclair}, \citenamefont {Peter}, \citenamefont {Ringeval},\ and\
  \citenamefont {Steer}}]{AuclairPeterRingevalSteer}%
  \BibitemOpen
  \bibfield  {author} {\bibinfo {author} {\bibfnamefont {P.}~\bibnamefont
  {Auclair}}, \bibinfo {author} {\bibfnamefont {P.}~\bibnamefont {Peter}},
  \bibinfo {author} {\bibfnamefont {C.}~\bibnamefont {Ringeval}},\ and\
  \bibinfo {author} {\bibfnamefont {D.}~\bibnamefont {Steer}},\ }\bibfield
  {title} {\bibinfo {title} {Irreducible cosmic production of relic vortons},\
  }\href {https://doi.org/10.1088/1475-7516/2021/03/098} {\bibfield  {journal}
  {\bibinfo  {journal} {Journal of Cosmology and Astroparticle Physics}\
  }\textbf {\bibinfo {volume} {2021}}\bibfield  {number} {\bibinfo  {number} {
  (03)},\ \bibinfo {pages} {098}},\ }\Eprint {https://arxiv.org/abs/2010.04620}
  {arXiv:2010.04620 [astro-ph.CO]} \BibitemShut {NoStop}%
\bibitem [{\citenamefont {Oliveira}\ \emph {et~al.}(2012)\citenamefont
  {Oliveira}, \citenamefont {Avgoustidis},\ and\ \citenamefont
  {Martins}}]{Oliveira:2012nj}%
  \BibitemOpen
  \bibfield  {author} {\bibinfo {author} {\bibfnamefont {M.~F.}\ \bibnamefont
  {Oliveira}}, \bibinfo {author} {\bibfnamefont {A.}~\bibnamefont
  {Avgoustidis}},\ and\ \bibinfo {author} {\bibfnamefont {C.~J. A.~P.}\
  \bibnamefont {Martins}},\ }\bibfield  {title} {\bibinfo {title} {{Cosmic
  string evolution with a conserved charge}},\ }\href
  {https://doi.org/10.1103/PhysRevD.85.083515} {\bibfield  {journal} {\bibinfo
  {journal} {Phys. Rev.}\ }\textbf {\bibinfo {volume} {D85}},\ \bibinfo {pages}
  {083515} (\bibinfo {year} {2012})},\ \Eprint
  {https://arxiv.org/abs/1201.5064} {arXiv:1201.5064 [hep-ph]} \BibitemShut
  {NoStop}%
\end{thebibliography}%

\end{document}